\documentclass[a4paper,fleqn,usenatbib,useAMS]{mnras}

\usepackage[T1]{fontenc}
\DeclareRobustCommand{\VAN}[3]{#2}
\let\VANthebibliography\thebibliography
\def\thebibliography{\DeclareRobustCommand{\VAN}[3]{##3}\VANthebibliography}
\usepackage[switch,columnwise]{lineno}

\usepackage{caption}
\usepackage{graphicx}	
\usepackage{amsmath}	
\usepackage{amssymb}	
\usepackage{multicol}        
\usepackage{bm}		
\usepackage{pdflscape}	

\usepackage{cleveref}

\newcommand{\nucal}{\textsc{nucal}}
\newcommand{\calamity}{\textsc{calamity}}

\title[Spectral Redundancy in 21\,cm Cosmology]{Spectral Redundancy for Calibrating Interferometers and Suppressing the Foreground Wedge in 21\,cm Cosmology}
\author[Cox et. al]{
Tyler A. Cox$^{1,2}$\thanks{E-mail: tyler.a.cox@berkeley.edu}
Aaron R. Parsons$^{1,2}$,
Joshua S. Dillon$^{1,2}$,
Aaron Ewall-Wice$^{1,2}$,
Robert Pascua$^{3}$,
\\$^{1}$Department of Astronomy, University of California, Berkeley, CA
\\$^{2}$Radio Astronomy Laboratory, University of California, Berkeley, CA
\\$^{3}$Department of Physics and Trottier Space Institute, McGill University, Montreal, QC, Canada
}
\date{Accepted 2024 June 20. Received 2024 June 18; in original form 2023 November 03
}
\pubyear{2024}
\graphicspath{{./}{figures/}}

\begin{document}
\label{firstpage}
\pagerange{\pageref{firstpage}--\pageref{lastpage}}
\maketitle

\begin{abstract}
Observations of 21\,cm line from neutral hydrogen promise to be an exciting new probe of astrophysics and cosmology during the Cosmic Dawn and through the Epoch of Reionization (EoR) to when dark energy accelerates the expansion of the Universe. At each of these epochs, separating bright foregrounds from the cosmological signal is a primary challenge that requires exquisite calibration. In this paper, we present a new calibration method called \nucal{} that extends redundant-baseline calibration, allowing spectral variation in antenna responses to be solved for by using correlations between visibilities measuring the same angular Fourier modes at different frequencies. By modeling the chromaticity of the beam-weighted sky with a tunable set of discrete prolate spheroidal sequences (DPSS), we develop a calibration loop that optimizes for spectrally smooth calibrated visibilities. Crucially, this technique does not require explicit models of the sky or the primary beam. 
With simulations that incorporate realistic source and beam chromaticity, we show that this method solves for unsmooth bandpass features, exposes narrowband interference systematics, and suppresses smooth-spectrum foregrounds below the level of 21\,cm reionization models, even within much of the so-called  ``wedge'' region where current foreground mitigation techniques struggle.
We show that this foreground subtraction can be performed with minimal cosmological signal loss for certain well-sampled angular Fourier modes,
making spectral-redundant calibration a promising technique for current and next-generation 21\,cm intensity mapping experiments.
\end{abstract}



\section{Introduction} \label{sec:intro}
Tomographic mapping of the redshifted hyperfine transition of neutral hydrogen (HI) holds great potential for studying the evolution of large-scale structure in the early universe. Successfully observing the cosmological 21\,cm signal will open a window for understanding the properties of the first stars and galaxies and constraining $\Lambda$CDM cosmology. For a review of 21\,cm cosmology, see, e.g., \cite{2006PhR...433..181F, 2010ARA&A..48..127M, 2012RPPh...75h6901P, 2020PASP..132f2001L}. Prior and ongoing experiments seeking to characterize patchy fluctuations in this 21\,cm signal include the Precision Array for Probing the Epoch of Reionization (PAPER; \citealt{2010AJ....139.1468P}), the Murchison Widefield Array (MWA;  \citealt{2013PASA...30....7T}), the Low-Frequency Array (LOFAR; \citealt{2013A&A...556A...2V}), the Hydrogen Epoch of Reionization Array (HERA; \citealt{2017PASP..129d5001D}), the Giant Metre Wave Radio Telescope (GMRT; \citealt{2013MNRAS.433..639P}), the Long Wavelength Array (LWA; \citealt{2019AJ....158...84E}), the Canadian Hydrogen Intensity Mapping Experiment (CHIME; \citealt{2014SPIE.9145E..22B}), and the Hydrogen Intensity and Real-time Analysis eXperiment (HIRAX; \citealt{2016SPIE.9906E..5XN}).

The principal challenge faced by experiments aiming to measure the high-redshift 21\,cm signal is to accurately separate the relatively weak cosmological signal from astrophysical foregrounds, particularly from continuum emission from our galaxy and other radio-bright galaxies, which are ${\sim}10^{4-5}$ times brighter. To address this challenge, experiments such as HERA and CHIME rely on the distinct spectral characteristics of foregrounds and the 21\,cm signal to separate the cosmological signal from the dominant foregrounds --- an approach known as foreground avoidance.

For redshifted line emission, spectral frequency corresponds to a line-of-sight spatial distance, making the Fourier dual of frequency a probe of line-of-sight modes (i.e.\ $k_\|$) in the 3D Fourier space of spatial fluctuations. HI is expected to exhibit significant spatial variation on cosmological scales in density, ionization state, and spin temperature, giving the cosmological 21\,cm signal structure at a variety of spectral scales \citep{2004ApJ...615....7M}. Foregrounds, on the other hand, are spectrally very smooth, and so occupy a relatively small number of low-order spectral modes.

In the absence of instrumental systematics, these differences in  spectral properties lead to a relatively clean separation between foreground contamination and higher-order spatial modes of the cosmological signal \citep{2010ARA&A..48..127M}.
However, this separation is degraded by the inherently chromatic nature of antenna beams and interferometric baselines, which modulate foreground power on spectral scales inversely proportional to the diameter or interferometric baseline length \citep{2012ApJ...756..165P}. The result is that foreground residuals tend to occupy a characteristic wedge shape in cylindrical $k$ space \citep{2010ApJ...724..526D,2012ApJ...756..165P, 2012ApJ...745..176V, 2013ApJ...768L..36P, 2013ApJ...776....6T, 2014PhRvD..90b3018L, 2014PhRvD..90b3019L}.

By avoiding modes within this wedge and focusing instead on the complementary EoR ``window'', detection of the cosmological 21\,cm signal remains possible, albeit with a reduced number of modes---and thus, sensitivity---for estimating the 21\,cm power spectrum. In practice, this picture is further complicated by spectral structure in signal chains, which convolutionally leak power from the foreground wedge into the EoR window, potentially overwhelming the cosmological signal. Preventing such spectral leakage requires precision calibration to invert any spectral structure that may have been imparted by the instrument. Given the high dynamic range between foregrounds and signal, chromatic effects larger than roughly one part in $10^5$ must be mitigated. 

Numerous approaches have been put forward to meet the challenge of accurately calibrating interferometers for 21\,cm cosmology (e.g.\ \citealt{2008arXiv0810.5751Y, 2010MNRAS.408.1029L, 2017arXiv170101860S, 2018ApJ...863..170L, 2018MNRAS.477.5670D, 2019ApJ...875...70B,2020MNRAS.499.5840D, 2020ApJ...890..122K}). These can be roughly organized into the categories of sky-based calibration and redundancy-based calibration. Sky-based calibration relies on accurate models of the sky and antenna beams to solve for per-antenna gains in visibility measurements, but this approach is susceptible to model error and incompleteness \citep{2016MNRAS.461.3135B, 2017MNRAS.470.1849E}. Redundancy-based calibration attempts to circumvent model incompleteness by using the fact that visibility measurements are expected to be identical for interferometric baselines with the same separation vector, provided antennas have identical beam patterns --- an assumption which can prove to be detrimental to the accuracy of derived calibration solutions if not realized in the field \citep{2019MNRAS.487..537O, 2019ApJ...875...70B}. In essence, redundancy-based calibration reduces the reliance on absolute knowledge of the sky and beam and focuses instead on symmetries internal to the antenna array.

In most implementations, both sky-based and redundant baseline calibration solve for antenna parameters independently versus frequency, relying on post hoc methods such as averaging \citep{2014MNRAS.445.1084Z}, gain smoothing \citep{2022ApJ...924...51A, 2023ApJ...945..124H}, or parametric fitting to limit the number of degrees of freedom in a calibration solution \citep{ 2015ApJ...809...61A, 2016MNRAS.460.4320E} in favor of calibrating all frequency channels in the data simultaneously. While this is not universally true (e.g., \citealt{2016arXiv160509219Y}), many calibration algorithms lack the capacity to calibrate all frequency channels simultaneously due to computing and memory requirements.

This is an unfortunate limitation, considering that the spectral axis provides the most powerful discriminant between astrophysical foregrounds and cosmological 21\,cm signal. As experiments move from setting upper limits to making detections of the 21\,cm power spectrum, there is a growing need for a rigorous accounting of the systematics that can arise from foregrounds interacting with low-level spectral structure in the instrument response. This motivates the need for calibration routines that can solve for the spectral structure while remaining robust to errors in sky and beam modeling.

In this paper, we introduce a new approach to calibrating radio interferometers for 21\,cm cosmology, which we call \emph{spectrally-redundant calibration} or \nucal, for short. This technique leverages the slow evolution of sky emission versus frequency to relate visibility measurements of the same angular Fourier modes at different frequencies. By extending the concept of redundant calibration along the frequency axis, this approach allows the calibration of arrays with arbitrary bandpass structure, while minimizing potential calibration errors introduced by incomplete sky models.

This paper is structured as follows: in \Cref{sec:cal_review}, we review the state of current calibration algorithms and motivate the need for new methods that optimize for the spectral smoothness of calibrated visibilities. In \Cref{sec:frc}, we introduce the concept of spectral redundancy and how it can be leveraged to calibrate 21\,cm arrays. We also introduce \nucal{}---our approach to redundant spectral calibration that enforces spectral and spatial smoothness in calibrated visibilities. In \Cref{sec:nucal}, we apply \nucal{} to simulated visibilities and show that we can faithfully solve for degenerate redundant-baseline calibration parameters and eliminate spurious structure into the calibration solutions. In \Cref{sec:wedge_recovery}, we demonstrate our ability to subtract foregrounds from simulated visibilities in a perfectly redundant array to recover the 21\,cm signal within the wedge. In \Cref{sec:discussion}, we discuss the assumptions made by spectrally redundant calibration and identify areas for further refinement. We conclude with a summary of our results in \Cref{sec:conclude}.

\section{Review of Calibration Techniques for 21\,cm Cosmology}\label{sec:cal_review}

In this section, we review the current status of 21\,cm calibration techniques and some of their limitations in order to motivate the need to improve direction-independent calibration methods by explicitly optimizing for spectral smoothness in calibrated visibilities. We begin with a brief review of the calibration problem.

\subsection{Calibration Problem}
Per-antenna gain calibration is the process of solving for a single complex number per antenna as a function of time and frequency. These numbers model signal chain effects and enter into the observed visibility of a baseline between antennas $i$ and $j$ as the true visibility multiplied by the frequency and time-dependent gain response of each antenna involved in the measurement,
\begin{equation}
    V^{\rm obs}_{ij}\left(\nu, t\right) = g_i\left(\nu, t\right) g_j^{*}\left(\nu, t\right) V^{\rm true}_{ij}\left(\nu, t\right) + n_{ij}\left(\nu, t\right)
    \label{eq:gain_cal}
\end{equation}
where $V_{ij}^{\rm obs}$ is the observed visibility between antennas $i$ and $j$, $g_i$ and $g_j$ are the complex gain of antennas $i$ and $j$, $V_{ij}^{\rm true}$ is the true visibility sampled by baseline, $b_{ij}$, and $n_{ij}$ is Gaussian distributed thermal noise on that measurement \citep{1996A&AS..117..137H, 2011A&A...527A.106S}. Additionally, most 21\,cm interferometers are dual-polarization instruments, which allow them to simultaneously measure electromagnetic signals from two orthogonal antenna polarizations. This requires that we also solve for frequency-dependent gains for each antenna polarization.

The process of correcting for these frequency, antenna, and polarization dependent gains is called \textit{direction-independent calibration}, which we will refer to simply as \textit{calibration} in this paper to distinguish it from \textit{direction-dependent calibration}, which accounts for the spatial response of each array element. Direction-independent calibration methods used for solving for antenna gains traditionally fall within one of two categories: sky-based calibration and redundant calibration. 

\subsection{Sky-Based Calibration}
One of the most common approaches to calibrating radio interferometers is through a method known as sky-based calibration. In sky-based calibration, detailed knowledge of the radio sky and antenna-beam pattern is employed to simulate a set of visibilities that ideally match the true, uncorrupted visibilities measured by the instrument. These simulated data products can then be used to set up an over-constrained system of equations to solve for antenna-dependent gain parameters. While this approach works quite well when the field of view is dominated by a single, bright, well-characterized source \citep{1965ApJ...142..122B}, arrays optimized for 21\,cm cosmology often have wide fields of view with elements that often have limited pointing ability, making it more challenging to build faithful sky and beam models for calibration. Additionally, diffuse Galactic emission can introduce calibration errors if not included in the model of the sky, particularly in short baselines which are more sensitive to diffuse foregrounds.

While sky maps have been slowly improving, existing experiments have not yet shown that the sky can be modeled to the precision necessary to calibrate visibilities to one part in $\sim$$10^5$. In particular, delivering foreground maps which accurately account for the spatial distribution, amplitude, and spectral properties of diffuse continuum emission from the galaxy and extra-galactic point sources to the precision required for 21\,cm cosmology proven very difficult. \cite{2016MNRAS.461.3135B} showed that unmodeled foreground point sources, even those below the confusion limit of current arrays, can introduce calibration errors which can limit the detection of the cosmological signal. These issues can be partially mitigated by down-weighting long baselines or only including relatively short, less-chromatic baselines when performing sky-based calibration \citep{2017MNRAS.470.1849E}. However, this can make calibration less accurate because this approach relies on baselines dominated by diffuse emission, which is rarely accurate to better than the percent level and is also highly polarized \citep{2016ApJ...830...38L}. In addition to making a measurement within the EoR window challenging, these errors make the level of foreground subtraction necessary for measuring cosmological modes within the foreground wedge impossible. 

Sky calibration is further complicated by the need to produce accurate models of the antenna primary beam, which typically have a complicated spatial response that is both polarization- and frequency-dependent. Significant work has gone into modeling instrument primary beam \citep{2016ApJ...831..196E, 2021ITAP...69.8143F} and validating these models by attempting to measure beams in-situ \citep{2012AJ....143...53P, 2016ApJ...826..199N, 2017PASP..129c5002J, 2018PASA...35...45L, 2018ExA....45....1D, 2018AJ....156...32E, 2018ExA....45..177P, 2020ApJ...897....5N}. However, state-of-the-art beam models are still insufficient to acquire sky-based calibration solutions that are accurate enough, or to do foreground subtraction to level necessary for 21\,cm cosmology. Even if the sky and beam are known to sufficient accuracy, ionospheric distortion can introduce frequency dependent errors in the position and intensities of known sources that can corrupt calibration solutions, making this method difficult to execute properly for 21\,cm experiments \citep{2017MNRAS.471.3974J, 2018MNRAS.478.1484G, 2021MNRAS.505.4775Y}.

\subsection{Redundant-Baseline Calibration}\label{subsec:redcal}

Another approach to calibrating an interferometer which largely skirts the issue of incomplete sky-models and imprecise knowledge of the beam is by calibrating an array assuming that antenna pairs which have the same baseline vector measure the same visibility value. Under this assumption, deviations from a shared measurement should be explainable by antenna-based gain terms that can be calibrated out of a raw visibility. This approach, known as \textit{redundant-baseline calibration} \citep{1992ExA.....2..203W, 2010MNRAS.408.1029L}, can be used to calibrate antenna gains in the case when repeated measurements of the same baseline vector are made without needing an a priori estimate of the true model visibilities. 

Redundant-baseline calibration seeks to solve for antenna-gain and visibilities parameters at each frequency by finding a solution to a system of equations of the form
\begin{equation}
V_{ij}\left(\nu, t\right) = g_i\left(\nu, t\right) g_j^*\left(\nu, t\right) V^{\rm sol}_{i-j}\left(\nu, t\right)
\end{equation}
which is done by minimizing $\chi^2$ written as,
\begin{equation}
    \chi^2 \left(\nu, t\right) = \sum_{i \neq j} \frac{\left|V^{\rm obs}_{ij}\left(\nu, t\right) - g_i\left(\nu, t\right) g_j^*\left(\nu, t\right) V^{\rm sol}_{i-j}\left(\nu, t\right)\right|^2}{\sigma_{ij}^2 \left(\nu, t\right)},
    \label{eq:redcal_chi2}
\end{equation}
where $V^{\rm sol}_{i-j}$ is the visibility solution for redundant baselines with the same baseline vector separation as $V_{ij}^{\rm obs}$ and $\sigma^2_{ij}$ is the noise variance on baseline $b_{ij}$. In the case where the number of unique baseline separations made by a perfectly redundant array is significantly less than the total number of measurements made (such as with HERA; \citealt{2016ApJ...826..181D}), a highly over-constrained system of equations can be set up to solve for the gains and model visibilities. In practice, no array is perfectly redundant and non-redundancies in the array due to small deviations in antenna position, beam shape, and beam pointing errors can all introduce chromatic gain errors which can complicate a 21\,cm measurement \citep{2019MNRAS.487..537O}. Nonetheless, redundant calibration remains an attractive approach since it is mostly free of assumptions about emission from the sky i.e. whether it is dominated by point sources, diffuse emission, or some combination of the two and does not initially require precision knowledge of the beam.

However, it is important to highlight that while redundant-baseline calibration reduces reliance on accurate models of the sky and beam, it ultimately still requires a sky-model to solve for a small number of remaining calibration degrees of freedom.
Regardless of how over-determined the system of equations used to minimize \Cref{eq:redcal_chi2}, the structure of $\chi^2$ guarantees that there will always be a few parameters as a function of frequency that cannot be solved by redundant-baseline calibration. These frequency-dependent parameters are degeneracies of this system of equations, which arise as a result of a set of transformations that one can apply to gains and visibilities that leave the value of $\chi^2$ unchanged \citep{2014MNRAS.445.1084Z, 2018MNRAS.477.5670D, 2020MNRAS.499.5840D}. These transformations that can be applied to the estimated gain parameters are an overall amplitude degeneracy,
\begin{align}
    g_i g^*_j V_{ij}^{\rm model} &\rightarrow \left(A g_i\right) \left(A g_j^*\right) \left(A^{-2} V_{ij}^{\rm model}\right) \nonumber \\
    &= g_i g_j^* V_{ij}^{\rm model},
    \label{eq:redcal_deg_amp}
\end{align}
and a phase gradient degeneracy,
\begin{align}
    g_i g^*_j V_{ij}^{\rm model} &\rightarrow \left(g_i e^{i \mathbf{\Phi \cdot r}_i}\right) \left(g^*_j e^{-i \mathbf{\Phi \cdot r}_j}\right) \left(V_{ij}^{\rm model} e^{-i \mathbf{\Phi \cdot b}_{ij}}\right) \nonumber \\
    &= g_i g^*_j e^{i \mathbf{\Phi \cdot b}_{ij}} V_{ij}^{\rm model} e^{-i \mathbf{\Phi \cdot b}_{ij}} \nonumber \\
    &= g_i g^*_j  V_{ij}^{\rm model}.
    \label{eq:redcal_deg_phs}
\end{align}
Here both $A$ and $\mathbf{\Phi}$ are real-valued frequency- and time-dependent quantities and $r_i$ is the position of antenna $i$. When each polarization is calibrated independently, there are three degeneracies\footnote{An additional overall phase degeneracy, where $g_{i} \rightarrow g_{j}e^{i\phi}$ does not generally need calibration because it leaves calibrated single-polarization visibilities unaffected.}  per frequency and polarization: an average gain amplitude and a phase gradient in both the North-South and East-West directions. 

After performing redundant-baseline calibration, an additional \textit{absolute calibration} must be performed to fill in this small set of degenerate calibration parameters per frequency, which amounts to setting the absolute flux scale and phase center for each frequency observed\citep{2010MNRAS.408.1029L, 2014MNRAS.445.1084Z, 2018MNRAS.477.5670D, 2019ApJ...875...70B, 2020ApJ...890..122K}. While this approach does significantly reduce the number of free parameters required to be solved for by a referencing a sky model, sky-model incompleteness errors can introduce calibration errors in redundant-calibration degeneracies which leak power into the EoR window that exceeds the 21\,cm signal \citep{2019ApJ...875...70B} if further measures are not taken to mitigate the effect.

\subsection{Hybrid Approaches}

It is worth noting that a number of hybrid calibration have been developed in recent years to address the issues detailed with both sky-based and redundant calibration. In \cite{2017arXiv170101860S}, a new approach \textsc{Corrcal} was presented which modeled covariance between baselines to bridge the gap between redundant-baseline calibration and sky-based calibration. \cite{2022MNRAS.510.1680G} was able to show that this approach led to moderate improvement in the calibration of PAPER data over redundant calibration alone. Other approaches, such as \textsc{BayesCal} \citep{2022MNRAS.517..910S, 2022MNRAS.517..935S} and Unified Calibration \citep{2021MNRAS.503.2457B}, have developed Bayesian frameworks to incorporate model uncertainty into calibration. They have shown a reduction in the spurious spectral structure of gain solutions compared to traditional sky-based and redundant calibration when estimating gain solutions in simulated data. The newly proposed Delay-Weighted Calibration \citep{2023ApJ...943..117B}, a variant of sky-based calibration, has also showed significant promise in reducing spurious spectral structure in gain solutions by down-weighting Fourier modes thought to have significant sky-model error.

\subsection{Mitigation of Spectral Structure in Calibrated Visibilities}
 
One of the primary challenges of 21\,cm cosmology is managing frequency-dependent calibration errors introduced when performing estimates of antenna-gain parameters. Assuming antenna gains are estimated accurately by some calibration method, one can simply divide the measured data by the estimated gains to recover the true sky signal. However, gain calibration is often performed inaccurately due to the issues mentioned above (i.e. inaccurate sky/beam models and non-redundancies). These errors often have fine-scale frequency structure which, due to the multiplicative nature of gain calibration (as in ~\Cref{eq:gain_cal}), are convolved with the true sky in Fourier space, leading to a leakage of foreground power into modes outside the wedge. Therefore, it is crucial that this calibration parameters be accurately estimated to prevent bright foreground from contaminating 21\,cm signal.

Many of the frequency dependent calibration errors introduced by both sky-based and redundant-baseline calibration have been mitigated by imposing the apriori assumption that the final calibrated gain parameters be smooth as a function of time and frequency. This assumption is generally imposed in a two-step process in which gains are estimated via sky-based or redundant calibration, then smoothed by fitting some set of slowly varying basis functions to the gain estimates \citep{2016MNRAS.461.3135B, 2018MNRAS.477.5670D, 2018MNRAS.478.1484G, 2019ApJ...887..141L, 2019AJ....158...84E}. Because instruments are designed to be stable in time and have smooth bandpass responses, imposing this a priori assumption is fairly reasonable in theory. It is therefore assumed that fine-frequency structure in gain solutions is more likely the result of calibration errors than true spectral structure in the instrumental response, and that it is therefore safer to ignore it than to introduce new chromaticity into calibrated visibilities. However, it remains to be seen whether these assumptions about smooth instrument responses hold in real-world antennas in the field. It is worth noting that there are exceptions to the approach of post-hoc smoothing calibration solutions, most notably \citet{2016arXiv160509219Y} which fits a parameterized version of antenna gains rather than smoothing post-calibration, but at the cost of greater computational expense and larger memory requirement.

These post-hoc approaches are suboptimal in the sense that they neglect a true accounting of what amount of true spectral structure may be found in the instrument bandpass or due to systematics such as cable reflections, which can be factored out of the data by modeling them as a gain-like term \citep{2019ApJ...884..105K, 2020ApJ...888...70K}. One can imagine that if the gains actually contain a significant amount of spectral structure, that post-hoc smoothing will remove those features from estimates of the gains, leading to structure which is not properly deconvolved from the calibrated visibilities potentially polluting the EoR window. For experiments to be successful in detecting the 21\,cm signal, calibration routines must be designed to handle imperfect knowledge of the sky and primary beam while remaining flexible enough to correct for true fine-scale frequency structure in instrumental gains. In the next section, we introduce our approach to calibrating fine-scale frequency structure using spectral redundancy---the repeated sampling of modes in the $uv$-plane by baselines at different frequencies.

\section{Spectral Redundancy}\label{sec:frc}
In this section, we explore the idea of spectral redundancy, an extension of traditional spatial redundancy to the spectral axis, which enforces consistency between measurements by comparing visibility measurements which sample the same $uv$-modes at different frequencies. We begin our discussion of spectral redundancy with a brief description of the visibility simulations used throughout the rest of this work (\Cref{sec:vissim}) and demonstrate the high degree of redundancy between baselines at frequencies which sample the same modes in the $uv$-plane (\Cref{subsec:frequency_redundancy}). We then discuss our approach to modeling spectrally redundant baselines by representing them as the sum of a flexible set of basis functions known as discrete prolate spheroidal sequences (DPSS). Finally, we revisit the antenna calibration problem and present our approach to calibrating data using spectral redundancy, \nucal{}, in \Cref{subsec:nucal}.

\subsection{Visibility Simulation Review} \label{sec:vissim}
\begin{figure*}
\centering
\includegraphics[width=.95\textwidth]{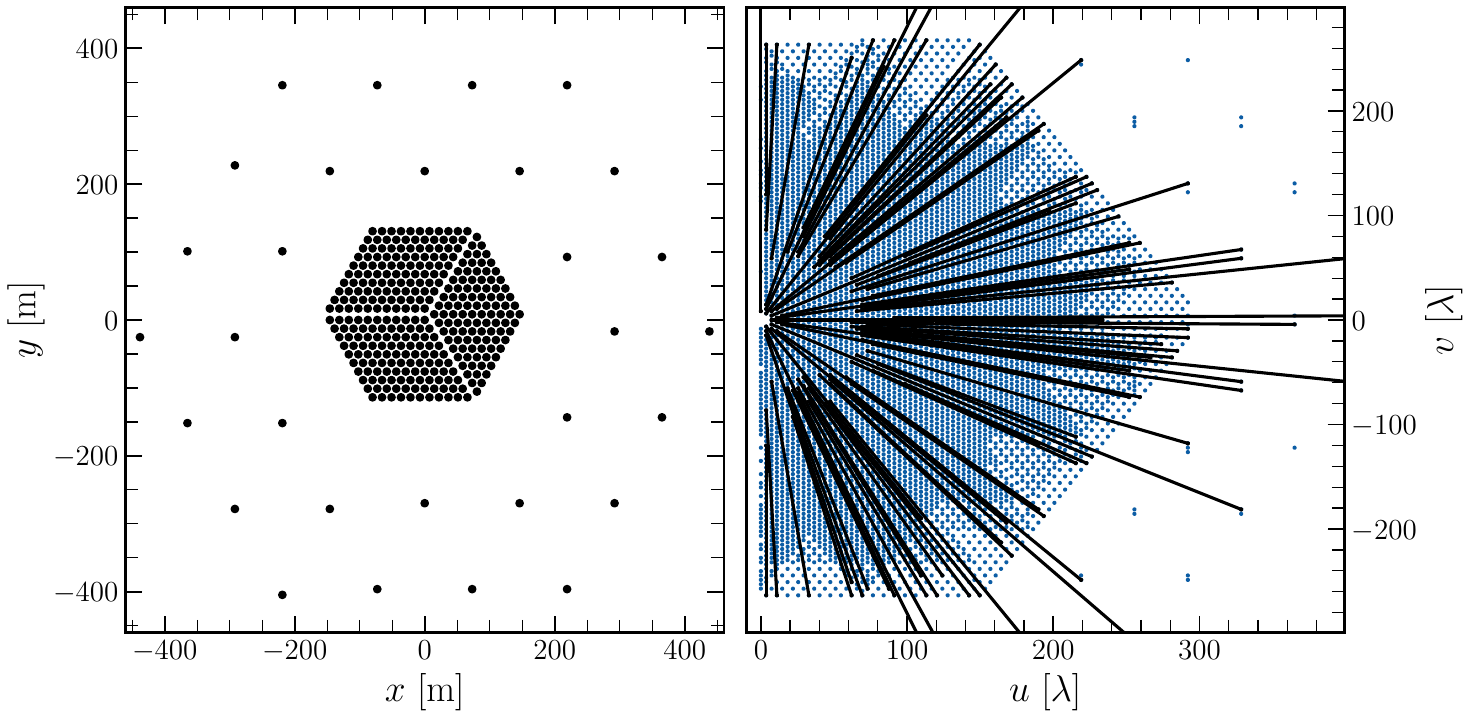}
 \caption{\textbf{Left Panel:} Array layout for the Hydrogen Epoch of Reionization Array (HERA). HERA was designed for redundant-baseline calibration, but its split-core configuration and 30 outrigger antennas effectively triple the number of unique baselines along any given heading in the $uv$ plane compared to a simple hexagonally-packed array \citep{2016ApJ...826..181D}. \textbf{Right Panel:} Corresponding samples in the $uv$-plane for the antenna positions shown in the left panel. Black lines overlaid on the $uv$-plane designate radial headings which contain baselines with significant overlap of same $uv$-modes at different frequencies (at least 15 
 unique baseline types within the same orientation, though we will use all orientations with at least 10 unique baseline types for the \nucal{} algorithm in \Cref{sec:frc}). 
 }
 \label{fig:antpos}
\end{figure*}

Before we begin with a description of spectral redundancy, we first give a brief description of the visibility simulations used in this work. In order to explore the extent to which measurements which made at the same angular Fourier mode at different frequencies are correlated, we use the antenna layout of the HERA array for our simulations. HERA was designed to have a high degree of spatial redundancy, but due to its regular layout of baselines in the same orientation, also has a significant amount of spectral redundancy as well, making it a good candidate for explorations of model real world arrays with spectral redundancy. HERA's layout and corresponding sampling of the $uv$-plane are shown in \Cref{fig:antpos}.

For our foreground model, we generate a set of $10^5$ point sources randomly distributed across the sky. The flux density of each point source follows the power law relation, 
\begin{equation}
    S\left(\nu\right) = f \left(\frac{\nu}{\nu_0}\right)^{\alpha}
    \label{eq:ps_vis}
\end{equation}
where $\alpha$ is the spectral index, $\nu_{0}$ the arbitrary pivot frequency, which we define to be $150$ MHz, and $f$ is the flux density of the point source at frequency, $\nu_{0}$. Our foreground sky model is defined as the sum of each of these randomly generated point sources,
\begin{equation}
    I\left(\hat{\mathbf{s}}, \nu\right) = \sum_i S_i\left(\nu\right) \delta\left(\hat{\mathbf{s}} - \hat{\mathbf{s}}_i\right)
\end{equation}
where $\delta$ is the Dirac-delta and $\hat{\mathbf{s}}_i$ is the unit vector describing the direction to that point. For each source in our simulation, we draw a random spectral index and flux density, where $\alpha$ is drawn from a uniform distribution ranging from $-1.5$ to $-0.5$, and the flux density is drawn from an exponential distribution with a mean of 300 mJy. In addition to a point source foreground model, we also simulate a flat power spectrum EoR by generating uncorrelated, Gaussian random HEALpix maps \citep{2005ApJ...622..759G} at each frequency. For these maps, we take the map size to be $\textsc{nside}=128$ ($N_{\rm pix} = 1.96 \times 10^5$) and with a variance of $25 \, {\rm mK^2}$, consistent with the expected fiducial brightness temperature of the signal at EoR redshifts \citep{2011MNRAS.411..955M}.

For our beam model, we choose an azimuthally-symmetric Airy function above the horizon, which roughly captures the spatial and spectral structure of a realistic instrument primary beam. This takes the form
\begin{equation}
    A\left(\nu, \hat{\mathbf{s}}\right) = \left(2\frac{J_1\left(\pi d_{\rm ant} \nu \sin\theta / c\right)}{\pi d_{\rm ant} \nu \sin\theta / c}\right)^2,
    \label{eq:airy_beam}
\end{equation}
where $\theta$ is the angle between zenith and the unit vector $\hat{\mathbf{s}}$, $J_1$ is the Bessel function of the first order and first kind, $d_{\rm ant}$ is the diameter of the antenna, which for HERA we take to be $d_{\rm ant} = {\rm 14 \ m} $, and $\lambda$ is the wavelength at which the beam is evaluated. While there are differences between the HERA beam and the Airy beam, they have similar spatial and spectral variability, making them a good candidate for realistic visibility simulations \citep{2016ApJ...826..199N}. 

With the beam and sky-model set, we use the visibility equation,
\begin{equation}
    V_{ij}(\nu) = \int d\Omega \ A\left(\nu, \hat{\mathbf{s}} \right) I\left(\nu, \hat{\mathbf{s}}\right) e^{-2 \pi i \nu \mathbf{b}_{ij} \cdot \hat{\mathbf{s}} / c}, \label{eq:RIME}
\end{equation}
to perform a discrete sum over each source in our sky-model weighting by the value of the Airy beam at the position of the source for each unique baseline orientation in the HERA array. Each visibility is simulated over the frequency range $46.9-234.9$ MHz with 1536, 122 kHz wide channels to match HERA's extended frequency range and channel width \citep{2021ITAP...69.8143F}.

In some figures later in this paper, we find it useful generate thermal noise for visibilities simulated to test the dynamic range capabilities of our technique. We generate this noise by randomly sampling from a complex Gaussian distribution with zero mean and a standard deviation set by the radiometer equation,
\begin{equation}
    \sigma_{ij} = \sqrt{\frac{V_{ii} V_{jj}}{\Delta t \Delta \nu}},
\end{equation}
where $V_{ii}$ is the autocorrelation amplitude of antenna $i$, $\Delta t$ is the integration time, which we set to HERA's integration time, $\Delta t = 10 \rm \ s$, and $\Delta \nu$ is the channel width.

\subsection{Spectrally Redundant Sampling of the $uv$-plane}\label{subsec:frequency_redundancy}

Traditional redundant-baseline calibration focuses on using similarities between baselines which share an orientation and physical length. However, within an array, there exist additional symmetries that can be leveraged to establish a concept of redundancy among measurements that are not conventionally considered redundant. Specifically, baselines that sample the same modes in the $uv$-plane at different frequencies---those sharing an orientation---may not yield identical visibilities values, but may still exhibit a form of redundancy in the sense that they are strongely correlated.

To understand how both spatial and spectral redundancy arise in interferometric measurements, we can generalize \Cref{eq:RIME} to express a visibility as a function of position in the $uv$ plane, $\mathbf{u} \equiv \nu\mathbf{b} / c$, equivalent to the baseline vector in wavelength units. It enters into the equation as the Fourier dual of the projected angular brightness on the sky,  
\begin{align}
    V\left(\mathbf{\mathbf{u}}, \nu\right) &= \int d\Omega \ A\left(\hat{\mathbf{s}}, \nu\right) I\left(\hat{\mathbf{s}}, \nu\right) \exp\left[-2 \pi i \mathbf{u} \cdot \hat{\mathbf{s}}\right].
    \label{eq:visibility_u}
\end{align}
From \Cref{eq:visibility_u}, it is clear that pairs of antennas that have the same separation vector $\mathbf{b}_{ij}$ should measure the same visibility, assuming that beam responses are identical. In practice, the redundancy that arises from the spatial arrangement of these antennas is broken by the unique signal delay and amplifier gain versus frequency of each antenna. However, as is mentioned in \Cref{subsec:redcal}, it is straightforward to set up a system of equations to solve for such per-antenna calibration terms in order to restore the redundancy of the underlying measurements.

Similarly, baselines which have dramatically different lengths may still be redundant in the sense that they measure very similar skies at different frequencies and thus be highly covariant. Consider the simple case of a linear array in which all the baseline vectors are oriented in the same direction. Let us also make the simplifying assumption that the sky and beam are achromatic, i.e\ $I\left(\hat{\mathbf{s}}, \nu\right) = I\left(\hat{\mathbf{s}}\right)$ and $A\left(\hat{\mathbf{s}}, \nu\right) = A\left(\hat{\mathbf{s}}\right)$. This allows to use \Cref{eq:visibility_u} to rewrite \Cref{eq:RIME} as
\begin{align}
    V_{ij}\left(\nu\right) &= \int d\Omega \ A\left(\hat{\mathbf{s}}\right) I\left(\hat{\mathbf{s}}\right) \exp\left[-2 \pi i \mathbf{u}_{ij} \cdot \hat{\mathbf{s}}\right] = V\left(\mathbf{u}_{ij}\right).
\end{align}
By ignoring the frequency-dependent of the sky and the beam in this toy example, we observe a given baseline's spectral characteristics are only dependent on the angular Fourier mode, $u$, that it measures. In this example, any two baselines in this linear array which sample the exact same angular Fourier modes ($\mathbf{u}_1 = \nu_1 \mathbf{b}_{1} / c = \nu_2 \mathbf{b}_2 / c = \mathbf{u}_2$), would produce identical visibility values despite potentially having vastly different physical lengths and making measurements at different frequency channels. These measurements could be compared and leveraged to constrain antenna gains, similarly to what is done in traditional spatial redundant-baseline calibration. 

\begin{figure*}
\centering
\includegraphics[width=.95\textwidth]{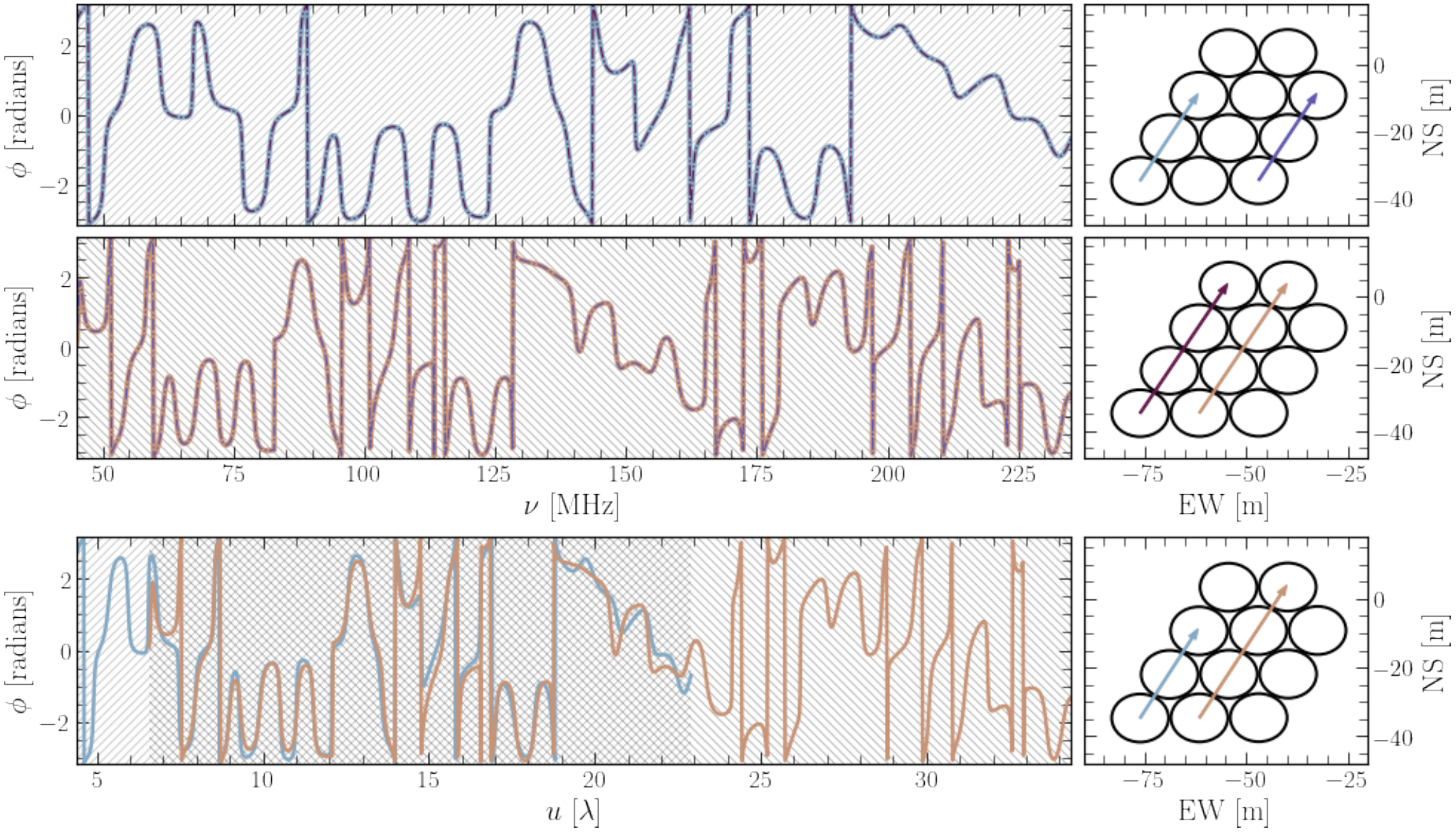}
 \caption{Traditional spatial redundancy involves comparing visibilities from antenna pairs separated by the same physical baseline vector, as we see in spectral phase plots of visibilities simulated with a chromatic beam and sky for 29\,m (top) and 44\,m (middle) baselines. Here, in these top two panels, the phase of each of the visibilities is plotted in the left column. In the right column, we show the corresponding antenna pairs and the baseline vectors connecting them. Because all antennas in the simulation share identical beams and the simulations are noiseless, the phases of the visibilities in the top rows align perfectly. In the lower panel, we overlay phase as a function of $u \equiv \nu |\mathbf{b}|/c$ for measured visibilities whose baselines have the same orientation but different lengths. Spectral redundancy is evidenced by the fact that both baselines see largely the same sky (at least in phase) at different frequencies. Their amplitudes will be quite different, since continuum foregrounds are might brighter at low frequency, but that spectral evolution is quite smooth at fixed $u$, as shown in \Cref{fig:tracks}. Note that though both baselines span the same range in $\nu$, they span different but overlapping ranges in $u$ (hashed regions).
 }
 \label{fig:redcal_vs_nucal}
\end{figure*}

While the sky and beam are not truly achromatic as described in this toy example, it is well known that the chromatic response of the foreground sky is quite smooth, as it is dominated by smooth spectrum processes such as galactic synchrotron emission. Likewise, antenna beams are designed to be as spectrally smooth as possible to prevent leakage of smooth spectrum foregrounds into the EoR window. Therefore, it is reasonable to assume that the sky-beam product $A\left(\mathbf{s}, \nu\right) I\left(\mathbf{s}, \nu\right)$ evolve smoothly with frequency, which would allow for a high degree of phase coherence between baselines which measure the same $uv$-modes. As shown in  \Cref{fig:redcal_vs_nucal}, baselines in the same orientation can in fact agree quite well in phase, provided the sky beam product evolve slowly as a function of frequency.

We can further explore the degree to which sample the same modes in the $uv$-plane are redundant by examining the cross-frequency, cross-correlation coefficient between pairs of baselines. Here, we define cross-frequency, cross-correlation coefficient as,
\begin{equation}
    \mathrm{R}_{ij} \left(\nu_1, \nu_2\right) = \frac{\mathbf{v}_{i, \nu_1} \cdot \mathbf{v}^*_{j, \nu_2}}{ \sqrt{\left(\mathbf{v}_{i, \nu_1} \cdot \mathbf{v}^*_{i, \nu_1}\right) \left(\mathbf{v}_{j, \nu_2} \cdot \mathbf{v}^*_{j, \nu_2}\right)}}
\end{equation}
where $\mathrm{R}$ is a matrix whose elements are the cross-correlation coefficients of pairs of frequencies sampled by baselines $i$ and $j$ and $\mathbf{v}_{i, \nu_1}$ a mean-subtracted visibility vector containing the visibility values from $N = 2000$ randomly-generated, point source simulations (as described in \Cref{sec:vissim}) for baseline $i$ at frequency $\nu_1$.

\begin{figure*}
\centering
\includegraphics[width=.99\textwidth]{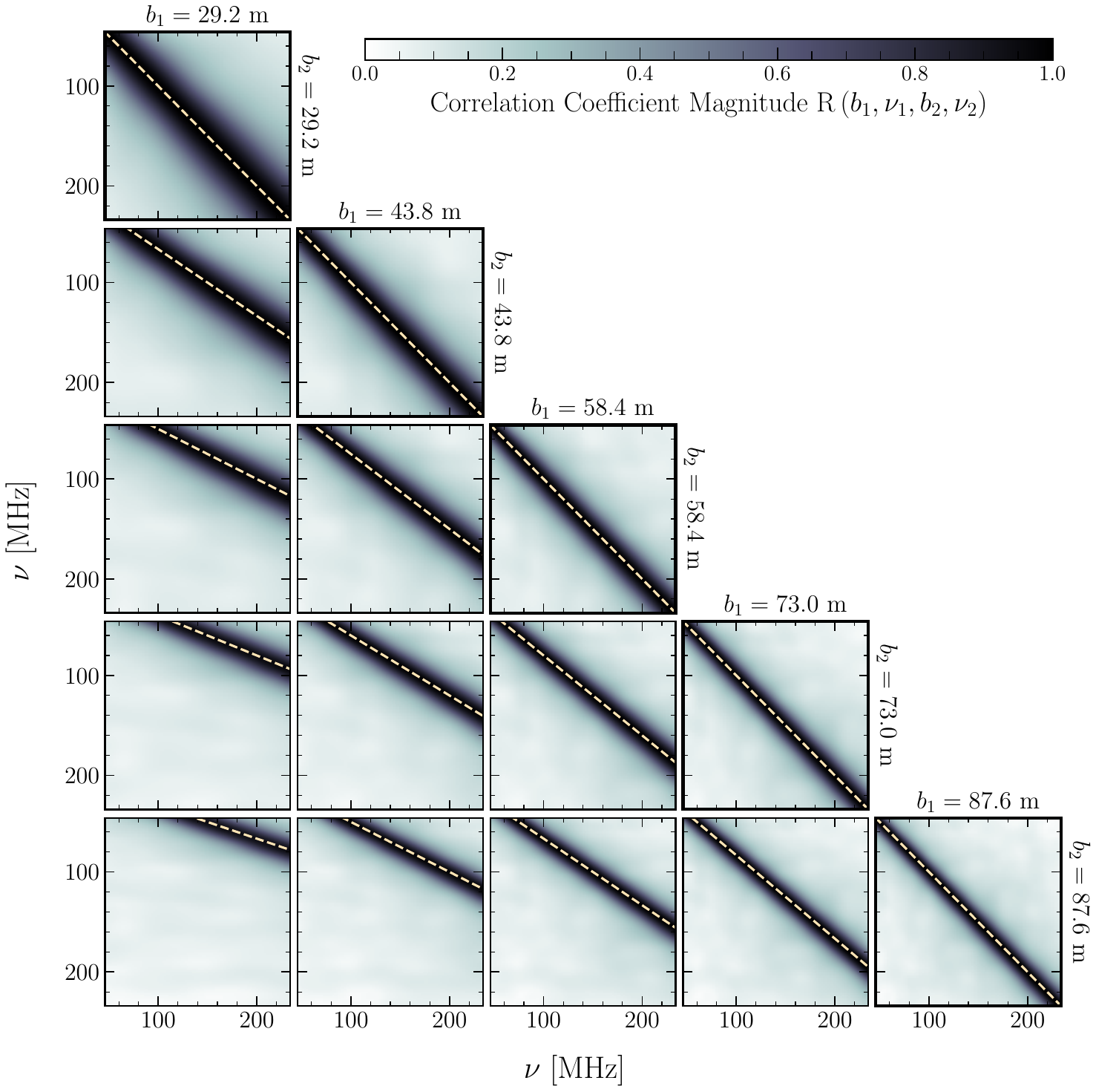}
 \caption{Cross-frequency correlation-coefficient matrices, $\mathrm{R}_{12} \left(\nu_1, \nu_2\right) = C_{12} / \sqrt{C_{11} C_{22}}$, computed by averaging over 1000 randomly-generated, noise-free point source foreground visibility simulations. Each correlation-coefficient matrix is computed between baselines in the same orientation, but with different lengths (except for the five panels on the diagonal, which show frequency-frequency correlations for the same baseline). The pale yellow dashed line in each plot follows, $\nu_2 = (b_1 / b_2) \nu_1$, and marks the location where baselines sample the same $uv$-mode, $\mathbf{u}_1 = \mathbf{u}_2$. Here, a clear correlation can be seen between baselines of different lengths, indicated by the near unity value of the cross-correlation coefficient, at frequencies where pairs of baselines redundantly sample the same angular Fourier modes. This strongly implies that different baselines at a fixed $\mathbf{u}$ are sensitive to the same fundamental sky-beam structure and, in some sense, redundant.
}
 \label{fig:correlation}
\end{figure*}

In \Cref{fig:correlation}, we show the cross-correlation coefficients for a sample of baselines within a group of baselines with the same azimuthal orientation in the HERA array. Along the main diagonal, we plot the cross-correlation matrices of each baseline  with itself, while the lower triangle features baselines crossed with another baseline of a different length.  As one might expect, baselines which have been cross-correlated with themselves are highly-correlated along their main diagonal. More interesting is the existence of a pronounced correlation between different baselines in the same orientation, particularly at frequencies where $\mathbf{u}_1 \approx \mathbf{u}_2$. This manifests in the near-unity cross-correlation coefficient in baselines of different lengths. 

\begin{figure*}
\centering
\includegraphics[width=.95\textwidth]{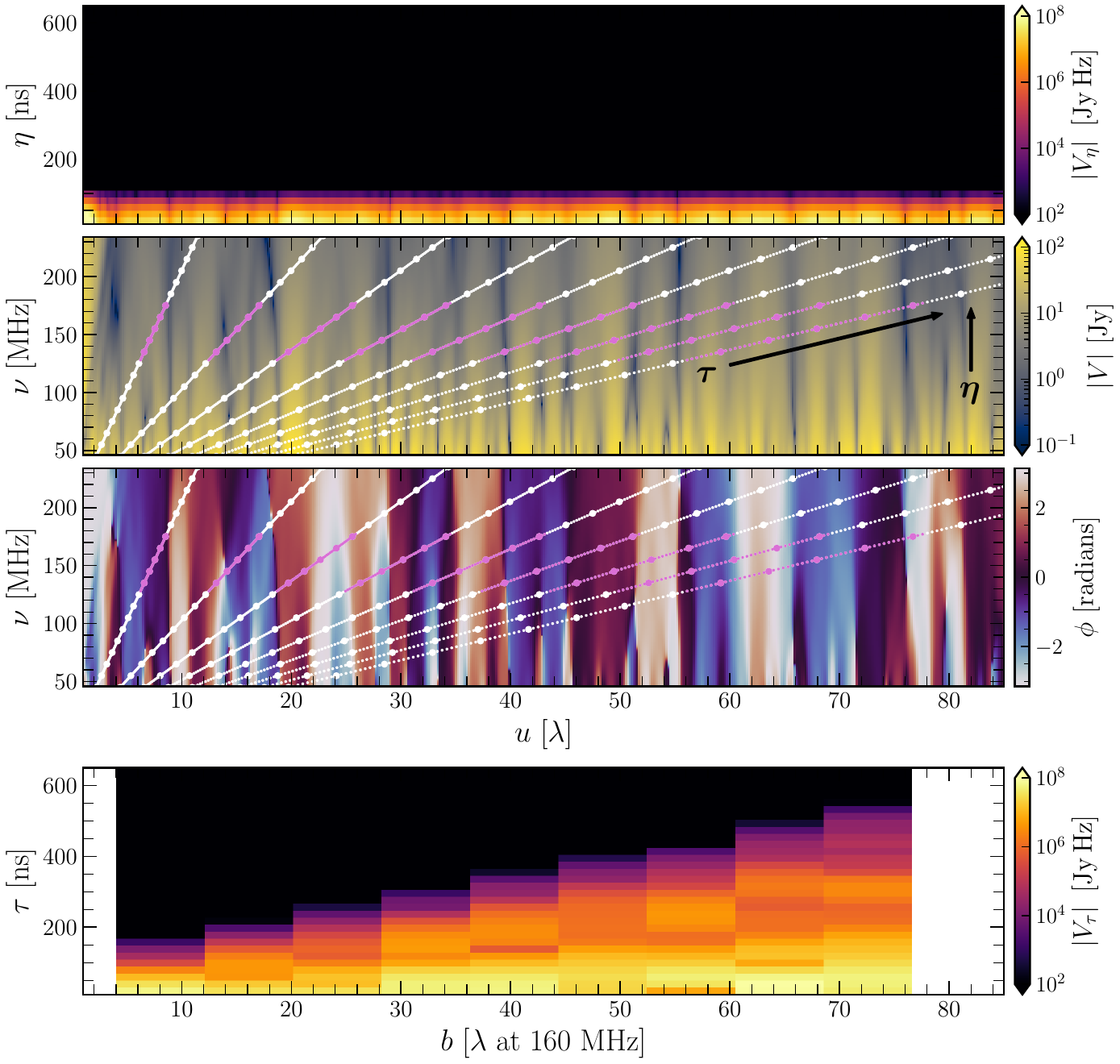}
 \caption{ A demonstration of the advantages of spectral redundant sampling of the $uv$-plane. In the center two panels, we plot the $u-\nu$ footprint of a group of 9 baselines oriented in the East-West direction within the HERA core in white over a simulation of visibility amplitudes and phases modeled with smooth spectrum foregrounds and an Airy beam, as a function of $u$ and $\nu$. Here, the large dots represent visibility measurements for every 100th channel and small dots represent every 10th channel (the true channel width of 122 kHz). All visibility measurements fall on these lines. In the bottom panel, we form the 2D power spectrum of the visibilities from the baselines plotted in the center two panels by taking the delay transform of each of the simulated visibilities individually over $130-170 \rm MHz$ (as shown by the magenta points). While the amplitude of the power spectrum is dominated by smooth spectrum sources, forming the power spectrum by taking the delay transform of the visibilities, $V_{\tau}$, mixes quickly varying $uv$-modes, which bleeds power, from the intrinsically foreground-dominated lower-order Fourier modes to the higher-order Fourier modes, leading to the wedge effect. In the top panel, we form the 2D power spectrum over the same range of frequencies, but instead take the Fourier transform of the visibility simulation perpendicular to the $uv$-axis, also known as the $\eta$-transform, $V_\eta$. Here, the $\eta$-transform highlights the localization of foreground power to low-$\eta$ modes which results from the Fourier transform being taken over a fixed angular scale. By comparing the measurements of baselines which redundantly sample the same modes in the $uv$-plane, one can infer the true frequency-structure of the foreground-beam product, despite the fact that the baselines themselves are highly chromatic as evidenced by the delay transform.
 }
 \label{fig:tracks}
\end{figure*}

This correlation means that different measurements are sensitive to the same underlying information. A question naturally arises: why would this new sort of redundancy be \emph{useful}? The answer comes from the fundamental difference between visibilities Fourier transformed along the frequency axis (or equivalently the line-of-sight axis) at fixed $\mathbf{u}$---the $\eta$ transform---and visibilities Fourier transformed along the same axis for a fixed baseline $\mathbf{b}$---the $\tau$ or delay transform. The delay transform is, in some sense, more natural, since it is performed directly on the measured visibilities. However, as we illustrate in \Cref{fig:tracks}, the $\eta$ transform is the space where foregrounds, even when multiplied by the beam, are the least chromatic.

In the center two panels \Cref{fig:tracks}, we plot the amplitude and phase of a simulated slice of the $uv$-plane, modeled with smooth spectrum foregrounds and a chromatic Airy beam as described in \Cref{sec:vissim}, as a function of $u$ and $\nu$. Overlaid on these panels are a series of large (small) white dotted lines, which represent every hundredth (tenth) frequency and the $u$ mode sampled there by a group of 9 baselines oriented in the East-West direction within the HERA array.

In the bottom panel, we construct the 2D power spectrum of the baselines depicted in the central two panels. We form this power spectrum by taking the Fourier transform over the frequency axis of each of the simulated visibilities individually (known as the delay spectrum) over the frequency range $130-170$ MHz, as indicated by the magenta data points in the second and third panels. While the amplitude of the power spectrum is dominated by smooth spectrum sources, the process of forming the power spectrum through the delay transform effectively mixes rapidly varying $uv$-modes. Consequently, this mixing results in the contamination of foreground power at lower delays, or $\tau$ (the Fourier dual of frequency in the delay regime) extending to higher delays. The extent of the power leakage is contingent on the range of $uv$-modes sampled by a given baseline, which is dependent on the baseline's physical length. This baseline-dependent mode-mixing effect is precisely what leads to the widely observed wedge effect that challenges the measurement of the 21\,cm signal. 

In the top panel, by contrast, we form the 2D power spectrum over the same frequency range, but this time applying a Fourier transform directly to the visibility simulation itself perpendicular to the $uv$-axis--$\eta$-transform. Unlike the delay transform, the $\eta$-transform does not introduce mixing between different $uv$-modes. Instead, it performs a Fourier transform along the frequency axis while keeping the $u$ coordinate fixed. Here, we see that the $\eta$-transform isolates beam-weighted foregrounds to many fewer modes than the delay-transform due to keeping the more chromatic axis of this plane $u$ fixed, leaving much of the Fourier space clean to measure the 21\,cm signal.  Put another way, the sky-beam product can accurately be described by many fewer parameters at fixed $\mathbf{u}$ than at fixed $\mathbf{b}$ for all but the shortest baselines.

While in practice we generally do not have access to all the required modes necessary to perform an $\eta$-transform and achieve this level of foreground isolation, this fundamental difference between the delay and $\eta$-transforms underscores the potential of redundant sampling of fixed $uv$-modes. If we can perform repeated sampling of fixed $u$-modes across frequency, as is done by comparing overlapping measurements between baselines in the same orientation which redundantly sample the same modes in the $uv$-plane, one can effectively differentiate between the intrinsic chromatic structure of a baseline that samples a broad range of $u$-modes and the true chromaticity of the beam-weighted sky. Leveraging the power of spectrally redundant sampling of the $uv$-plane for precise estimates of the sky chromaticity holds great promise for a number of applications. In particular, it proves to be a powerful tool in calibration applications, solving for antenna gains which cause visibilities to deviate from the expected chromaticity at fixed $u$-modes, and for improved foreground suppression. We explore both applications later in this paper.

\subsection{Constructing a Model for Spectrally-Redundant Visibilities with DPSS}

One important distinction between spatial redundancy and spectral redundancy and is that while two or more spatially redundant baselines can be directly compared when calibrating, it is much less straightforward to compare measurements that sample the same modes in the $uv$-plane with spectral redundancy---a point we highlight in \Cref{fig:redcal_vs_nucal} and \Cref{fig:tracks}. In traditional spatial redundancy, one expects that two baselines with the same separation vector should be redundant up to the per-antenna, per-frequency complex gains, assuming perfect redundancy in the array (i.e. identical beams and perfect antenna placement). Because each baseline makes the same measurements at the same frequency channels, comparing these measurements to enforce spatial-redundancy between baselines is trivial. 

However, comparing measurements in an attempt to enforce the less-than-perfect spectral redundancy between two or more baselines that sample the same $uv$-mode is not as simple. Simply put, while the true visibility evolves smoothly with frequency at fixed $u$, it still does evolve and that evolution must be parameterized and modeled. Further, we must deal with the complexity introduced by the fact that baselines in the same orientation that cross a given mode in the $uv$-plane at different frequencies rarely sample the \emph{exact} same modes. Both challenges must be tackled in order to compare measurements across baselines.

One prior approach to incorporating spectral redundancy into calibration tried to solve this problem by statistically modeling the correlations between baselines in order to solve for the models of the visibilties which were spectrally redundant. In \citealt{2021MNRAS.503.2457B}, the authors introduced a technique for incorporating spectral redundancy into a unified Bayesian framework, citing private communication with some of us during the development of this work. Their technique forward-modeled the cross-frequency covariance between baselines, incorporating information about the sky and beam model into the estimate of the covariance matrices. These covariance matrices were then used in an optimization loop which penalized non-redundant features between pairs of visibilities when solving for a set of spectral redundant visibility models. This technique has the benefit of modeling the redundancy between baselines which may not strictly sample the same modes in the $uv$-plane, but nearby modes. This approach makes the technique very flexible, but extremely computationally challenging for arrays with wide-bandwidths and many frequency channels to take advantage of spectral redundancy. Additionally, this approach requires accurate knowledge of the sky and primary beam to product estimates of the cross-baseline frequency-frequency covariance matrices used in the optimization loop. 

Ideally, we would like our technique to make assumptions which are informed by realistic properties of the sky and the beam, but not \emph{require} precise knowledge of either to perform accurate calibration and modeling of the spectrally redundant visibilities. Additionally, we would like to strictly enforce that our modeling procedure explicitly limits the modeled foreground visibilities to the level of spectral structure we expect. To achieve these requirements, we choose to model our spectrally redundant visibilities as a linear combination of a finite number of highly-efficient basis functions known as discrete prolate spheroidal sequences (DPSS) or the Slepian basis. These basis functions have the ideal property of having their power being maximally concentrated within a contiguous region of Fourier space with a half-bandwidth of $\eta_{w}$ while also being orthonormal, making them a powerful set of basis functions for modeling band-limited signals. 

In recent years, the DPSS basis has gained popularity in the 21\,cm community as an efficient basis for per-baseline modeling of foregrounds \citep{2021MNRAS.500.5195E}, in-painting missing data in channels flagged for radio frequency interference (RFI)  \citep{2023MNRAS.520.5552P}, and for performing gain calibration \citep{2022ApJ...938..151E}. As a mathematical basis, DPSS modes span the space of variations subtended by interferometric measurements of foregrounds while providing a rigorous formulation for limiting the scale of modeled structure---particularly in applications with irregular sampling such as in-painting or modeling flagged data---using pseudo-inverses of an analytically calculable covariance matrix.

The DPSS basis functions are defined with respect to some finite band within which we want to maximally concentrate the power of that function. The spectral concentration ratio for some function $\varphi$ is the Fourier domain of $\nu$, $\eta$, is defined as follows,
\begin{equation}
 \mu = \int_{-\eta_{w}}^{\eta_{w}} |\varphi\left(\eta\right)|^2 d\eta \Big/ \int_{-\eta_{\rm max}}^{\eta_{\rm max}} |\varphi\left(\eta\right)|^2 d\eta,
 \label{eq:spec_concentration}
\end{equation}
where $\eta_{w}$ is the half-width in $\eta$ that we wish to maximally concentrate the power of $\varphi$ and $\eta_{\rm max}$ is the Nyquist limit for the discrete grid of points over which the discrete band in $\nu$ is defined. By maximizing this quantity, one obtains the first DPSS mode for a given $\eta_{w}$. The full set of DPSS basis functions is formed for this $\eta_{\rm max}$ by repeatedly solving for the function that maximizes $\mu$ and is also orthogonal to all prior solutions to \Cref{eq:spec_concentration}. This procedure provides a basis which is complete within the bounds $-\eta_{w} < \eta < \eta_{w}$ and also maximally concentrates its power within these bounds. \cite{1978ATTTJ..57.1371S} showed that this optimization problem could be summarized by solving the eigenvalue problem,
\begin{equation}
    \sum^{N_d - 1}_{n=0} \varphi_n\left(N_d, \mathcal{W}\right) \frac{\sin \left[2 \pi \mathcal{W} \left(m - n\right)\right]}{\pi \left(m - n\right)} = \varphi_n\left(N_d, \mathcal{W}\right) \lambda_n\left(N_d, \mathcal{W}\right),
    \label{eq:eigenval_problem}
\end{equation}
where $\mathcal{W} = \eta_{w} \Delta \nu$ is a parameter which sets the filter half-width of the filter in Fourier space, $\varphi_n\left(\nu\right)$ is the $n$th DPSS function, $N_d$ is the number of discrete points where $\varphi$ is evaluated, and $\lambda_n$ is the corresponding eigenvalue of the $n$th DPSS function. 

While the number of functions in a complete set of DPSS basis functions is equal to the number of samples in the basis, $N_d$, typically only a few basis vectors are actually needed to represent a given band-limited signal to high precision if it is spectrally smooth. Which basis functions are used for modeling a given band-limited signal is typically determined by the eigenvalue spectrum of the DPSS eigenfunctions. The eigenvalue spectrum of DPSS modes has a characteristic step shape where significant eigenvalues have a value near 1 and whose power is primarily concentrated within $\eta_w$ and insignificant eigenvalues have a value near 0 and have significant power leakage beyond $|\eta| > \eta_w$, with a few eigenvalues with intermediate values ($1 > \lambda > 0$). The approximate number of eigenmodes with near unity eigenvalue values asymptotes to $\sim$$2 B \mathcal{W}$ when $N \rightarrow \infty$ \citep{1978ATTTJ..57.1371S, 2020arXiv200600427K}. The shape of this eigenvalues spectrum allows us to select a finite number of basis functions to fit or filter data by taking only those with eigenvalues greater than some eigenvalue cutoff $\epsilon$. We perform this fitting and/or filtering of data presumed to be band-limited using the linear least-squares approach laid out in the DAYENUREST formalism in \citep{2021MNRAS.500.5195E}. In this paper, we take $\epsilon = 10^{-12}$.

For this paper, we utilize the DPSS basis to modeling groups of baselines within the same orientation as the outer product of two sets of DPSS vectors in order to model the spatial and the spectral axes. Namely, 
\begin{equation}\label{eq:NUCALMODEL}
    V^{\rm model}_{i-j}\left(\nu\right) = \sum_{m} \sum_{n} a_{mn} \varphi_{m}\left(\|\mathbf{u}_{ij}\|, \ell_{\rm max}\right) \varphi_{n}\left(\nu, \eta_{\rm max}\right)
\end{equation}
where $\varphi_m$ and $\varphi_n$ are the sets of basis functions that model the $uv$-axis and frequency axis respectively and $\ell_{\rm w}$ and $\eta_{\rm w}$ are parameters which set the Fourier half-width of the $uv$-axis and frequency axis respectively. By taking the outer product of these set of basis functions, we effectively form a new set of basis functions whose power is concentrated in a rectangular region of 2D Fourier space within the Fourier domains of $u$ and $\nu$ ($\ell$ and $\eta$ respectively). Representing baselines in this way allows us to precisely limit the model to the level of structure that we expect to be present in each axis of a set of spectrally-redundant visibilities. For the spatial axis, we restrict the DPSS basis to evolve no faster than $-1 < \ell < 1$ as limited by the horizon (i.e. half-wavelength scales in $u$). For the spectral axis, we inform of choice of DPSS based on stringent limits set by antenna chromaticity \citep{2016ApJ...825....9T, 2016ApJ...826..199N, 2016ApJ...831..196E, 2018ExA....45..177P, 2020ApJ...897....5N, 2021MNRAS.500.1232F, 2021ITAP...69.8143F} and foreground smoothness \citep{2000ApJ...530..133T, 2006ApJ...650..529W, 2008MNRAS.388..247D, 2012MNRAS.419.3491L, 2017MNRAS.464.3486Z}.  For this paper, we take the Fourier half-width of the spectral axis to be $\eta_{w} = 25 \ \rm ns$.

We compute the DPSS basis functions along the frequency axis by using the \textsc{scipy} implementation \citep{2020SciPy-NMeth}, which employs the Lanczos algorithm to solve the eigenvalue problem (\Cref{eq:eigenval_problem}). This approach necessitates that the samples along the axis where the basis functions are generated be uniformly-spaced. Given that each baseline within the HERA array sample the same frequencies and the frequency channels are sampled uniformly, this condition is satisfied.

In contrast, the spatial axis lacks a uniformly spaced grid that is consistent across all baselines in the group that would allow us to use the \textsc{scipy} implementation. This discrepancy arises from the fact that the spacing of discrete samples along the $uv$-axis is determined by the length of the baseline (i.e $\Delta u = \Delta \nu \mathbf{b} / c$). Therefore, each baseline in the spectrally redundant group samples the $uv$-axis non-uniformly, which requires that we generate values from the Slepian basis that exactly correspond with the location each baseline samples in the $uv$-plane. To satisfy this requirement, we instead generate samples from the continuous extension of the discrete prolate spheroidal sequence---the prolate spheroidal wave functions.\footnote{Note that discrete prolate spheroidal wave functions are discrete sequences sampled from the continuous prolate spheroidal wave functions. If the two sets of functions had the same Fourier half-width support and sampled the same points, the values of the functions would be identical.}

We generate these functions by utilizing an implementation that solves the continuous eigenvalue problem described by:
\begin{equation}
    \int \frac{\sin \left[2 \pi \eta_w \left(\nu - \nu^{\prime}\right)\right]}{\pi (\nu - \nu^{\prime})} \varphi_n(\nu^{\prime}, \eta_w) d\nu^{\prime} = \lambda_{n} \varphi_{n}(\nu, \eta_w).
\end{equation}
This method uses a normalized Legendre polynomial expansion to sample from the prolate spheroidal wave functions at the exact values of $u$ each baseline measures. Further details regarding the process for generating these samples can be found in \cite{Moore_Cada_2004}.

\begin{figure*}
\centering
\includegraphics[width=.99\textwidth]{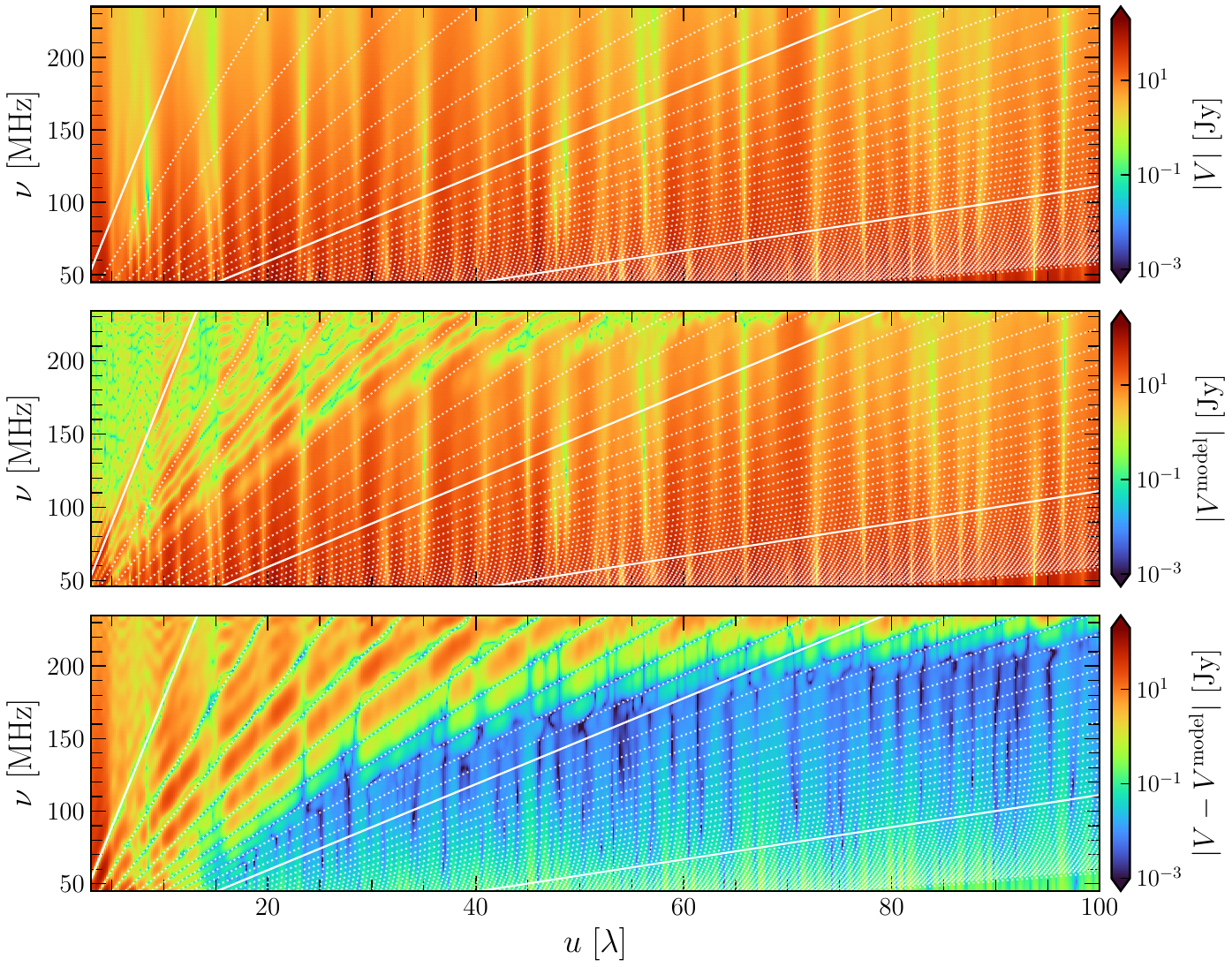}
\caption{Here we demonstrate the DPSS model's ability to accurately fit the $u$-$\nu$ plane with a relatively restricted set of basis vectors. In the top panel, we show true visibility without noise, evaluated using \Cref{eq:RIME}, for one of the best-sampled spectrally-redundant groups in the array. This is done not just at the points actually measured by the array (shown in white), but across the entire plane. We then show in the middle panel our model (\Cref{eq:NUCALMODEL}), which is fit using only the discrete set of visibility samples. Finally, we show the magnitude of the difference in the bottom panel. The bold white lines indicate the baselines we show the relative error of the model in \Cref{fig:nucal_modeling_residuals}. When the frequency density of visibility samples is high enough---i.e., where the spectral redundancy is large enough---the model is a good fit and can even accurately interpolate between measured points (blue regions). Where the spectral redundancy is lower, such as in the upper left, the fit is underdetermined and only the actual sampled visibilities are well-modeled. This figure can give us a sense of where we expect foreground modeling and subtraction with spectral redundancy to be effective and where we expect it to be lossy and to remove cosmological signal in the wedge---a point we return to in \Cref{sec:wedge_recovery}.
}
\label{fig:nucal_modeling}
\end{figure*}

To demonstrate that this basis and parameterization is sufficiently flexible to model astrophysical foregrounds as observed by a realistic beam, we explore how well we can model the entire $u$-$\nu$ plane (similar to \Cref{fig:tracks}). The fit, which we show in \Cref{fig:nucal_modeling} for one of the best-sampled spectrally-redundant groups in the array, is constrained only by the discretely sampled visibilities in the spectrally redundant group. Despite that, not only are the observable visibilities fit well, but much of the unobserved parts of the plane are fit well too. This strongly implies that with sufficient spectral redundancy, the DPSS model is overdetermined by the visibilities and that that extra information could be used to subtract foregrounds within the wedge. Of course, this raises a question of signal loss, which we return to in \Cref{sec:wedge_recovery}.

\begin{figure}
\centering
\includegraphics[width=.99\columnwidth]{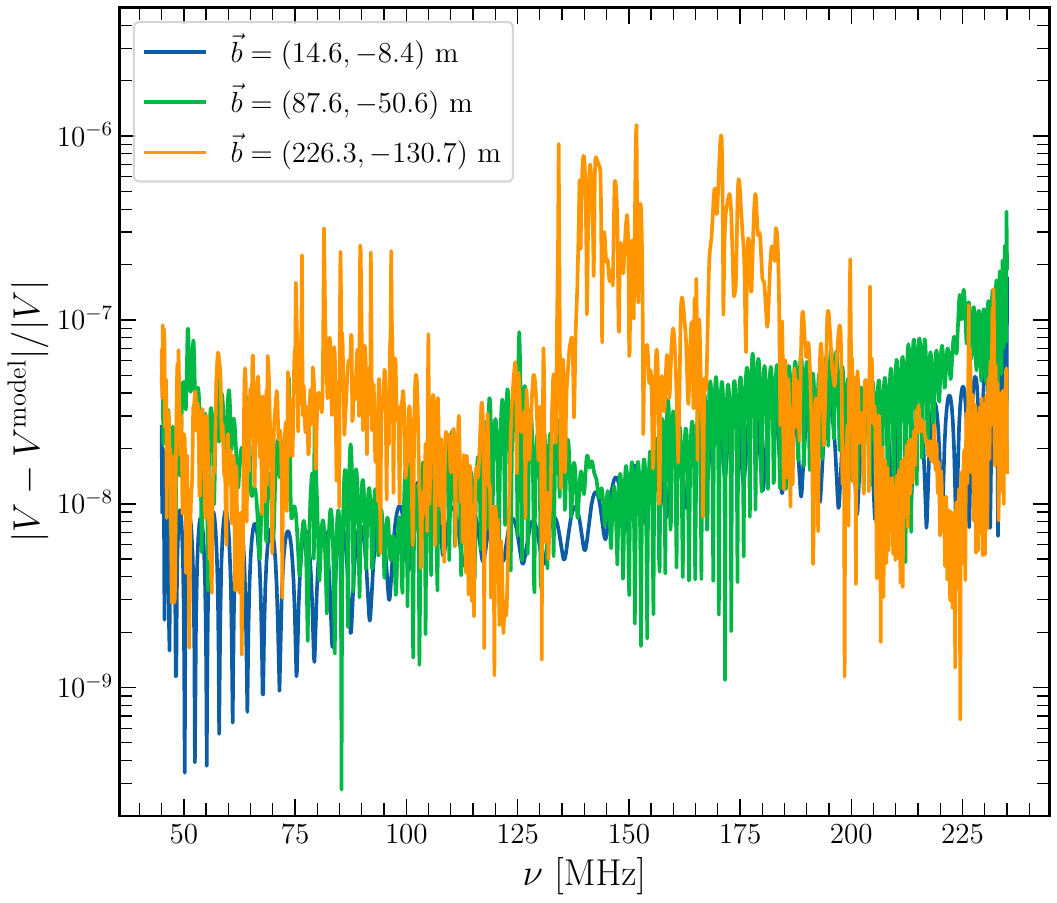}
\caption{To more quantitatively assess the fidelity of the \nucal{} model fit in reproducing the simulated visibilities, we plot the relative error of the model shown in \Cref{fig:nucal_modeling} evaluated for the three baselines highlighted as bold white lines. Regardless of the level of spectral redundancy at the $u$ modes probed by these baselines, the model is able to fit a foreground-only simulation to a very high dynamic range---better than 1 part in $10^6$. This is despite the fact that the \nucal{} is restricted to only modeling slow frequency evolution at fixed $u$. This demonstrates the flexibility of the model, though we expect that the level of spectral redundancy will still matter when assessing cosmological signal loss (see \Cref{sec:wedge_recovery}).
}
\label{fig:nucal_modeling_residuals}
\end{figure}

Regardless, by examining the quality of the fit to the actual measurable visibilities, which we show in \Cref{fig:nucal_modeling_residuals}, we can see that the model and its limited range of basis vectors is sufficient to model foregrounds to better than 1 part in $10^6$. At least to the level of realism reflected by our simulations, this is more than enough dynamic range to precisely model and remove foregrounds to below the expected EoR level. This is very promising, though there are certainly real-world complexities that any application of this technique would have to address, as we discuss in \Cref{subsec:realworld}.

\subsection{The Robustness of the DPSS Model to Spectrally Localized Features}
\label{subsec:rfi_w_model}
One downside to incorporating cross-frequency, cross-baseline information into a model fit, as is described in the previous subsections, is the potential for that fit to be corrupted by features in the visibilties which are compact in frequency and large in amplitude. These features, which violate the assumptions of spectral redundancy, can lead to model errors which can ripple across the frequency band of one or more baselines. One such spectrally localized feature commonly found in visibility data is radio frequency interference (RFI). 

\begin{figure}
\includegraphics[width=.95\columnwidth]{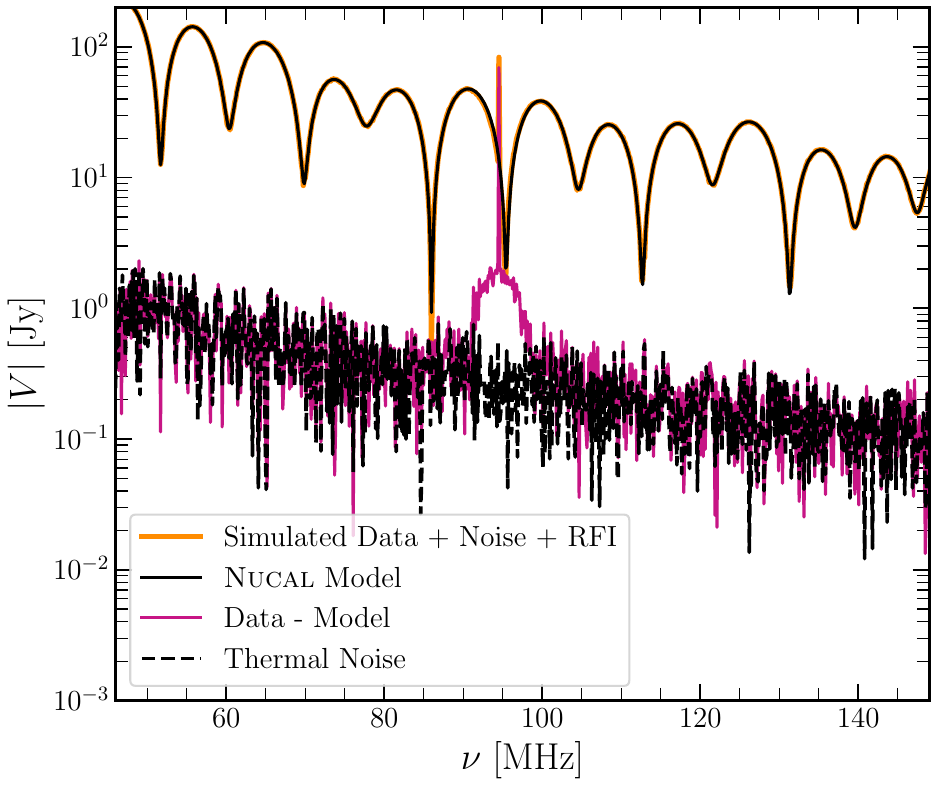}
\caption{Demonstration of the robustness of the DPSS basis being used to model a group of spectrally-redundant baselines to real-world effects. This modeling basis is restricted to modeling structure which is smooth as a function of frequency at a fixed $u$, and rejects shapes in visibilities that are spectrally compact. To demonstrate this ability, we simulate radio frequency interference (RFI) in one frequency channel across all baselines in the spectrally redundant group (58 baselines with the same orientation), and fit our model to the simulations with noise and RFI added. Leveraging spectral redundancy, we find that the DPSS basis functions are able to produce a model (solid black line) of smooth spectrum foreground emission of one of those baselines (in this case 9.3\,m east, 12.0\,m south) to the level of the thermal noise (dashed black line) while mostly rejecting the RFI spike, which can be seen in the data-model residuals (purple). The residuals show remarkable agreement with the thermal noise simulated across most of the band, demonstrating the ability of the DPSS basis to model and remove foregrounds. In particular, this technique may be adept at identifying low-amplitude narrowband RFI, which is difficult to locate using traditional methods and has the potential to introduce high-delay structure which can swamp the cosmological signal. A more thorough investigation of that potential is left for future work.
}
\label{fig:rfi_blip}
\end{figure}

In \Cref{fig:rfi_blip}, we demonstrate this effect by fitting the DPSS basis described in the previous section to the most spectrally redundant group of visibilities within the HERA array --- the same group used in \Cref{fig:nucal_modeling} and \Cref{fig:nucal_modeling_residuals}. Prior to performing the fit, we simulate unflagged RFI in the measurement set by adding a $50$ Jy RFI blip to all the simulated visibilities at a single frequency channel near $100 \ {\rm MHz}$. We then perform a linear-least squares fit of the DPSS basis to the visibilities with the RFI added and plot the simulated visibilties, the DPSS model, and residuals. From the figure, we observe that fitting the DPSS basis to a set of baselines corrupted by RFI does in-fact lead to errors which impact nearby channels, as can be seen in the residuals. However, thanks to the stringent constraints imposed by spectral redundancy and the DPSS basis, our set of basis function are still able to model and remove foreground emission down to the thermal noise floor over the majority of the band. 

While these types of errors are concerning if we wish to produce high-fidelity models of foreground emission over a wide-range of frequencies, there are techniques which can minimize the impact of the spectrally localized features, such as detecting and flagging the worst offending sources of RFI before fitting and implementing procedures which identify RFI-impacted channels during the fitting process. This potential pitfall may also to prove a strength of spectral-redundancy---if certain channels exhibit a particularly strong discrepancy between data and model, that maybe be evidence for narrowband RFI that should be flagged before re-running the fit.

\subsection{Spectrally Redundant Calibration (\nucal{})}\label{subsec:nucal}

For highly spectrally-redundant baseline groups (see \Cref{fig:antpos}), the parameters describing the smooth foregrounds in \Cref{eq:NUCALMODEL} are overdetermined. This means that the data can simultaneously constrain the foreground model parameters and nuisance parameters like calibration gains. 
Just as in redundant-baseline calibration, our goal when we perform that fit is to minimize $\chi^2$, which represents the noise-weighed sum of the difference between our measured and visibilities modeled with DPSS modes in \Cref{eq:NUCALMODEL}. Here, we generalize \Cref{eq:redcal_chi2},
\begin{equation}
    \chi^2 = \sum_{ij\nu} \frac{\left|V^{\rm obs}_{ij}\left(\nu\right) - G_{ij}\left(\nu\right) V_{i-j}^{\rm model}\left(\mathbf{u}_{ij}, \nu\right)\right|^2}{\sigma_{ij}^2\left(\nu\right)}
    \label{eq:chi2_nucal},
\end{equation}
where we seek to solve for both for our visibility model and antenna gains. Here $\sigma^2_{ij}$ is a weighting function that incorporates thermal noise variance and any flags, which are treated as having infinite variance, and $G_{ij}$ are a set of gain parameters to be solved for in calibration. While in redundant-baseline calibration, $V^{\rm sol}_{i-j}$ is allowed to have arbitrary spectral structure considering that each frequency is solved for independently, \nucal{} restricts $V^{\rm model}\left(\mathbf{u}_{ij}, \nu\right)$ to be the sum of a limited number modes at fixed $u$ that vary smoothly in frequency. 

By using multiple orientations within the $uv$-plane which have baselines with significant overlap and modeling each with the DPSS basis, \Cref{eq:chi2_nucal}, the simultaneous solution of  $G_{ij}(\nu)$  and $V^{\rm model}\left(\mathbf{u}_{ij}, \nu\right)$ becomes even more over-determined---depending on the degrees of freedom we assign to our model for $G_{ij}(\nu)$. Restricting the visibility model to a set of relatively smooth DPSS vectors in $u$ and $\nu$ allows this calibration to minimize spectral structure in calibrated visibilities and attempts to put as much of the observed spectral structure in uncalibrated (or partially-calibrated) visibilities into the gains. Where traditional spatially-redundant calibration uses internal symmetries in an array to calibrate antenna gains to one another at each frequency independently, \nucal{} uses analogous symmetries in the spectral response to explicitly calibrate the array along the frequency axis. This allows spectrally redundant baselines to be brought into alignment with a physically-motivated model for the angular arrangement and chromatic variation of the bright foregrounds that dominate their measurements. 

This is one of the primary benefits of spectral redundancy: it provides us a new-axis for enforcing self-consistency between visibilities. Although spectral redundancy does not require that two different baselines precisely measure identical visibility values at the same angular Fourier modes across various frequencies, it does impose stringent constraints on the extent of spectral variations at a constant value of $u$ across baselines of vastly different lengths. Because we anticipate these measurements to exhibit gradual changes with frequency at fixed angular scales, deviations from the anticipated level of spectral smoothness observed in the raw visibilities can often be ascribed to non-smooth signal chain effects. These deviations are subsequently accounted for as direction-independent antenna gain terms during the fitting process.

While \nucal{} can in principle be applied to solve for per-antenna gains, for this paper we restrict \nucal{} to only work within the degenerate space of spatially-redundant baseline calibration. \nucal{} is a somewhat natural solution to fixing the degeneracies of redundant-baseline calibration without introducing spectral structure into the degenerate parameters, as many arrays which have leaned heavily on spatial redundancy, such as HERA and PAPER, are also capable of being calibrated using spectral redundancy. The procedure for solving for these degeneracies is quite straightforward within the framework of \nucal. We restrict $G_{ij}(\nu)$ to take the form
\begin{equation}
    G_{ij}\left(\nu\right) = A\left(\nu\right) \exp\left[{i{\mathbf \Phi}\left(\nu\right) \cdot {\mathbf b}_{ij}}\right], \label{eq:nucal_gain}
\end{equation}
where $A$ and $\Phi$ are the frequency-dependent amplitude and phase tip-tilts degeneracies described in \Cref{subsec:redcal}. In this way, \nucal{} replaces (or at least refines) the absolute calibration step that must follow after redundant-baseline calibration. More importantly, it does so in a way that optimizes for the spectral smoothness of the calibrated visibilities and thus reduces the impact of gain errors due to sky-model incompleteness.

Because \nucal{} relies on fitting many spectrally-redundant groups simultaneously to constrain these array-wide degeneracy parameters, and because adding gains makes the fit no longer linear, the computational challenge is dramatically increased compared to fitting a DPSS model to a single spectrally-redundant group. Therefore, we implement the calibration routine described above as a first-order gradient descent optimization in \textsc{Jax} and \textsc{optax} \citep{jax2018github} to take advantage of the built-in auto-differentiation, just-in-time compilation, XLA acceleration, and library of popular optimizers. As written, the gradient descent algorithm makes the problem of calibrating every frequency channel simultaneously tractable, but only converges to the local minima closest to our initial guess of the model parameters. Therefore, \nucal{} is best applied after an initial sky-based absolute calibration has already been performed.

It is worth noting that this technique bears several similarities to the recently proposed calibration technique, CALAMITY \citep{2022ApJ...938..151E}, which also utilizes the DPSS basis to model foregrounds and solves for antenna gains by minimizing a loss function using first-order gradient, but does so in a way that does not incorporate spectrally redundant information into the fit. For a review of the similarities and differences between these two approaches, see \Cref{appendix:calamity}.

\subsubsection{\nucal{} Degenerate Parameters}\label{subsubsec:nucal_degens}
Like spatially-redundant calibration, spectrally-redundant calibration is not a full solution of all the calibration degrees of freedom. Given that \nucal{} calibrates on the basis of internal consistency between baselines that redundantly sample the same modes in the $uv$-plane at different frequencies, there are a set of transformations that can be applied to either the gains or visibility models that ultimately leave the value of $\chi^2$ in \Cref{eq:chi2_nucal} unchanged.
For example, one can take $A \rightarrow 2 A$ from \Cref{eq:nucal_gain} and $a_{mn} \rightarrow \frac{1}{2} a_{mn}$ from \Cref{eq:NUCALMODEL} without modifying the product $G_{ij} V_{i-j}^\text{model}$ in \Cref{eq:chi2_nucal}. The precise number and shape of these degeneracies depends on one's array layout and DPSS parameterization.

However, because we enforce that the visibility model is spectrally smooth at a fixed $uv$-mode across baselines in a spectrally redundant group, the number of \nucal{} degeneracies must be dramatically reduced from the number of degeneracies that must be fixed by traditional absolute calibration after spatially-redundant calibration. There were at least three physically important numbers per frequency per polarization. Here, we expect only a handful of such numbers over the whole band. This largely eliminates the need for sky-based absolute calibration to faithfully calibrate fine-scale spectral structure, and thus reduces the impact sky or beam modeling errors can have on the final calibration solution. 

In practice, one can use \nucal{} by starting with a best-attempt at sky-based absolute calibration and letting the degeneracies simply be left unmodified by the gradient descent solver, which only takes steps in non-degenerate directions. Because the visibility model is restricted to one consistent with smooth spectrum foregrounds, the dynamic range of foreground subtraction (see \Cref{sec:wedge_recovery}) is not affected by absolute calibration errors in the degenerate subspace of \nucal{}. This of course assumes that perfectly calibrated foregrounds are well-captured by the parameterized \nucal{} model. While one can still get errors in the degenerate subspace that affect the magnitude of the recovered EoR signal as a function of redshift, these sorts of errors do not risk mixing the foreground signal into uncontaminated modes, as is the case in the standard direction-independent calibration problem.

\section{Demonstration of Spectrally-Redundant Calibration (\nucal{}) on Simulated Data} \label{sec:nucal}
In this section, we test the performance of \nucal{} on a set of simulated visibilities which contain realistic spectral variation and calibration errors.  We begin with a description of the gains applied to the simulated data to produce a set of ``uncalibrated'' data.

\subsection{Setting Up the Problem}

For the simulations used in this section, we assume that the data have already been redundantly-calibrated and that the only remaining calibration required is the removal of the degeneracies of spatially-redundant calibration. This is done to reduce the computational cost of performing a per-antenna gain calibration, but is still a scenario of practical relevance given that the arrays designed to perform spatial redundant calibration also generally have some degree of spectral redundancy (e.g.\ PAPER and HERA). Here, we simulate the need for such a calibration by moving our model visibilities within the degenerate parameter space of redundant-baseline calibration. To highlight \nucal's ability to accurately account for arbitrary amounts of spectral structure in an instrument bandpass, we simulate the degenerate amplitude (\Cref{eq:redcal_deg_amp}) and tip-tilt parameters (\Cref{eq:redcal_deg_phs}) to vary rapidly from channel-to-channel by drawing the amplitude from a normal distribution with a mean of 1 and standard deviation of 0.01 and the tip-tilt degeneracy from a normal distribution with a mean of 0 radians per meter and standard deviation of 0.01 radians per meter for both the north-south and east-west parameters. 

We then use \nucal{} to calibrate this corrupted dataset by selecting baselines within orientations which have greater than 10 unique baseline types to participate in the estimation of the gains. For each unique orientation, we construct a foreground model, allowing each set of foreground components to vary independently of all other orientations included in the calibration and restrict the set of basis functions modeling the spatial axis to only model out $u = 100 \lambda$. We then execute the first order gradient descent to refine our initial estimates of the foreground model components and the redundant-baseline degeneracies. 
The final product of the \nucal{} implementation produces a DPSS-based model of the sky, $V^{\rm model}_{i-j}$, a set of calibrated visibilities, $V_{ij}$, and an estimate of the spatially redundant degeneracies, $A\left(\nu\right)$ and $\mathbf{\Phi}\left(\nu\right)$. Since our primary goal is to test the dynamic range of this calibration procedure, our simulations are noise-free. See \Cref{sec:vissim} for details on visibility simulations and \Cref{tab:simulation-details} for a summary of simulation and calibration parameters.

\begin{table*}

\label{tab:simulation-details}
\begin{tabular}{ll}
\hline
\multicolumn{1}{|l|}{Parameter}                & \multicolumn{1}{l|}{Value}                                       \\ \hline
Instrument Layout                              & HERA-350 (Hexagonal Core and Outrigger Antennas; Figure 1)       \\
Frequency Range                                & 46.9$-$234.4\,MHz                                                 \\
Frequency Resolution ($\Delta \nu)$            & $122$\,kHz                                                        \\
Sky Model                                      & $10^5$ Point Sources (\Cref{eq:ps_vis}) + Flat-Spectrum Power Spectrum \\
Beam Model                                     & Airy Beam (\Cref{eq:airy_beam})                                           \\
Spatial Axis DPSS Half-Width Parameter & 0.5$\lambda$\\
Frequency Axis DPSS Half-Width Parameter ($\eta_{w}$) & $25$\,ns                                                          \\
Minimum Number of Unique Baselines per Group   & 10                                                               \\
Model Constraints (Data Points / Parameters) & 11.94 \\
\hline
\end{tabular}
\caption{Visibility Simulation and Calibration Details}
\end{table*}

\subsection{Demonstration of \nucal{}'s Dynamic Range in Fourier Space}\label{subsec:dspec}
After \nucal{}, we perform a delay transform, in which we substitute a Fourier transform along the frequency axis of a visibility for a line-of-sight Fourier transform \citep{2012ApJ...756..165P, 2014PhRvD..90b3018L}, 
\begin{equation}
    \widetilde{V}\left(\tau\right) = \int d\nu \ V\left(\nu\right) W\left(\nu\right) e^{-2 \pi i \tau \nu}
    \label{eq:delay_transform}
\end{equation}
where $W\left(\nu\right)$ is a frequency taper function applied to the data prior to Fourier transforming, which we take to be a Blackman-Harris function \citep{1978IEEEP..66...51H}.
We perform this delay-transform on both the \nucal{}-calibrated visibilities, and the uncalibrated visibilities and the visibilities of the foregrounds and EoR separately.

\begin{figure*}
\includegraphics[width=.99\textwidth]{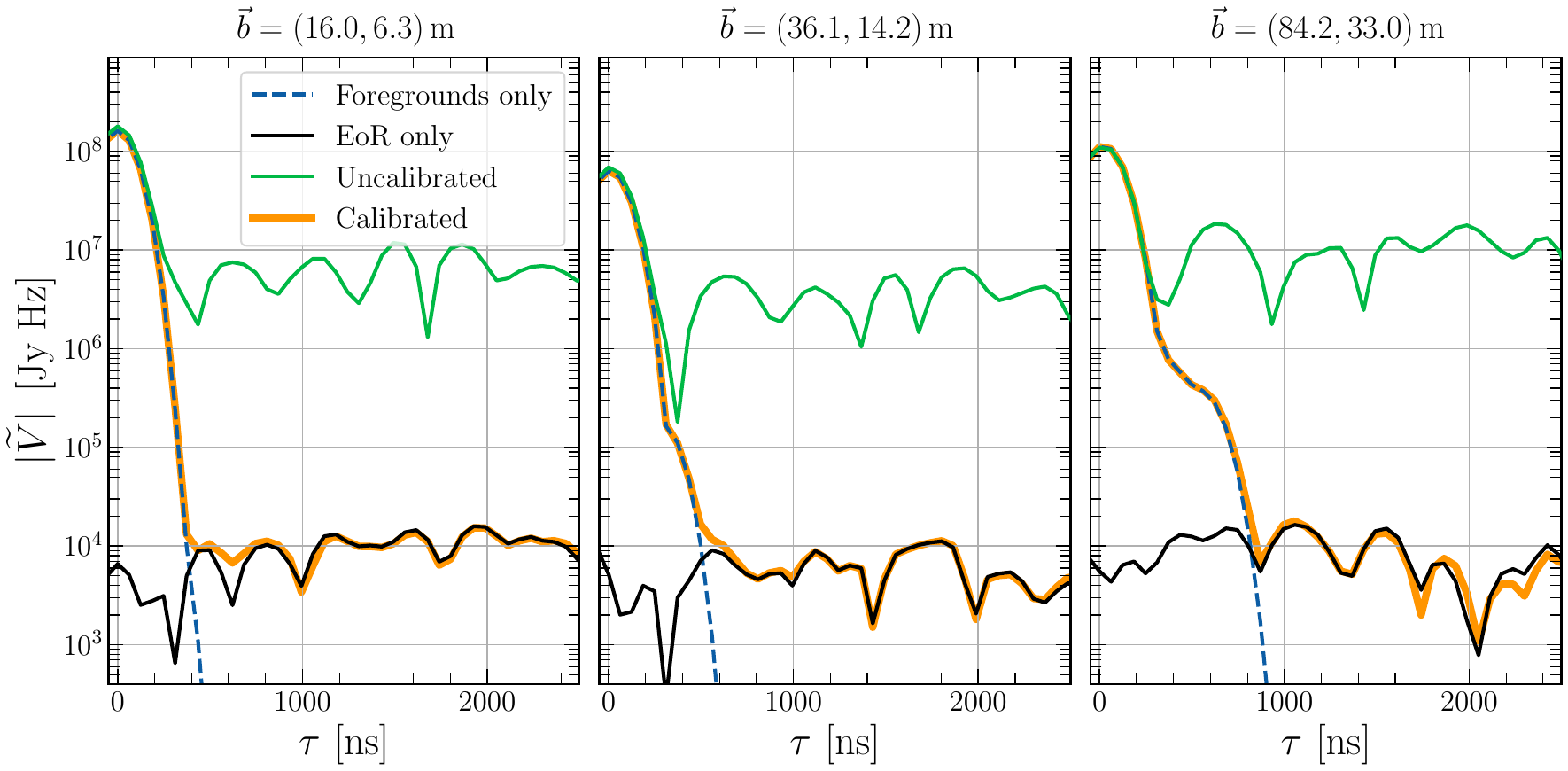}
\caption{Demonstration of \nucal{}'s ability to calibrate out arbitrary spectral structure introduced by an instrument for three different baselines within the same spectrally redundant group. Plotted in green is a noise-free but intentionally uncalibrated simulated visibility, with gains only in the degenerate subspace of redundant-baseline calibration. We also plot the power spectrum of noise-free simulated foreground visibilities (blue dashed), the EoR power spectrum (black), and the visibility after spectrally redundant calibration (orange). We find that \nucal{}, by restricting the level of spectral structure allowed in the calibrated visibility, correctly removes the spatially-redundant degenerate parameters in our simulated visibilities, leaving calibrated visibilities consistent with the EoR power spectrum outside the wedge.}
\label{fig:delay_spec}
\end{figure*}

In \Cref{fig:delay_spec}, we inspect the delay-transformed visibilities associated with three baselines that belong to the same spectrally redundant group. The primary goal of this analysis is to assess the effectiveness of \nucal{} in mitigating spurious spectral structure imparted by a highly chromatic bandpass. For each of the baselines plotted, we find that \nucal{} is effectively able to solve for and thus calibrate out the redundant-calibration degenerate parameters. More importantly, \nucal{} is able to converge to a solution of the degeneracies at a precision necessary to recover the 21\,cm signal outside the wedge for each of the baselines plotted. This demonstrates \nucal{}'s utility for dramatically reducing the complexity of absolute calibration after redundant-baseline calibration and for isolating the result from sky and beam modeling errors that would impart spurious spectral structure to contaminate the EoR window.

\section{Spectral Redundancy as a Method of Foreground Modeling and Subtraction}
\label{sec:wedge_recovery}

Potentially the most interesting application of spectral redundancy is the production of incredibly accurate sky models. The model of the sky that we fit during spectrally redundant calibration, can also be used as a spatio-spectral filter of beam-weighted smooth-spectrum sky emission. When the sampling of $uv$ modes is sufficiently dense along a given orientation, this model constrains foreground emission at a given frequency using information from a wide range of other frequencies---much wider than the coherence scale of cosmological 21\,cm emission. This means that, for certain orientations and frequency ranges, this technique has the ability to remove foregrounds while retaining 21\,cm emission within the wedge, even with a realistic level of sky and beam chromaticity. Access to these modes would greatly increase the sensitivity of 21\,cm experiments as a significant number of SNR modes lost to foreground contamination, particularly interferometers with many long baselines which must sacrifice more sensitivity than compact arrays to estimate the delay spectrum using only modes outside the wedge (the so-called ``foreground avoidance'' strategy). 

In this section, we attempt to use spectral redundancy to excise foregrounds from a set of simulated visibilities while still leaving modes within the wedge. For ease of analysis, we assume that our simulated visibilties are perfectly calibrated so that the effectiveness of this basis for removing foregrounds while leaving 21\,cm signal recoverable can be evaluated without the complication of instrumental systematics. In order to examine our ability to recover cosmological signal within the wedge for existing instruments, use the single most spectrally-redundant group of visibilities within the HERA array. With this set of uncorrupted simulated visibilities, we use the set of DPSS modeling vectors to perform a least-squares fit to the simulated data allowing the set of filters modeling the $u$-axis the freedom to model structure that varies at scales of half a wavelength and the set of filters which model the spectral axis to model variation with a Fourier half-width in $\eta$ of $\eta_{\rm max} = 25 \ \rm ns$ as was done in \Cref{sec:frc} and \Cref{sec:nucal}.

After computing this model, we form an estimate of the power spectrum by forming delay-spectra for each baseline within the spectrally redundant group for each of the simulated data products (simulated 21\,cm signal, foregrounds, 21\,cm signal and foregrounds, and the visibilities filtered by subtracting the \nucal{}) using the delay-transform where the visibilities are multiplied by a Blackman-Harris window function (\Cref{eq:delay_transform}). We take the delay-spectrum over the frequency band, 60$-$70\,MHz, where more baselines in the group have more significant spectrally-redundancy. It should be noted that while the power spectrum is estimated over a small sub-band, the \nucal{} model is fit over the full frequency range of 46.9$-$234.4\,MHz. We then estimate the 2D power spectrum by squaring the delay spectra and scaling to cosmological power spectrum units using,
\begin{equation}
    \hat{P}_{ij}\left(\textbf{k}_\perp, k_\parallel\right) = \frac{X^2 Y}{\Omega_{pp} B} \left|\widetilde{V}_{ij}\left( \tau\right)\right|^2,
\end{equation}
\citep{2014ApJ...788..106P} where $\hat{P}_{ij}$ is our estimate of the power spectrum as measured by baseline, $b_{ij}$, $\Omega_{pp}$ is the solid angle integral of the primary beam squared, $X$ and $Y$ are linear scaling factors which map comoving distance to angular separation and radial comoving distances to the frequency interval $\Delta \nu$,
\begin{align}
    X\left(z\right) &= \frac{c \left(1 + z\right)^2}{\nu_{21} H(z)} \\
    Y\left(z\right) &= \frac{c}{H_0 \nu_{21}} \frac{\left(1 + z\right)^2}{E\left(z\right)}
\end{align}
and $B$ is the bandwidth over which the delay transform was taken \citep{2006PhR...433..181F, 2012ApJ...756..165P}. We also transform our baseline and delay coordinates into cosmological spatial modes perpendicular, $\mathbf{k}_{\perp} = 2 \pi \mathbf{b}_{ij} / X$, and parallel to the line of sight, $k_\parallel = 2 \pi \tau / Y$ \citep{2012ApJ...753...81P}. For all results shown,  we adopt a $\Lambda$CDM cosmology derived from the Planck 2018 analysis when computing power spectra and cosmological coordinates \citep{2020A&A...641A...6P}.

\begin{figure*}
\centering
\includegraphics[width=.99\textwidth]{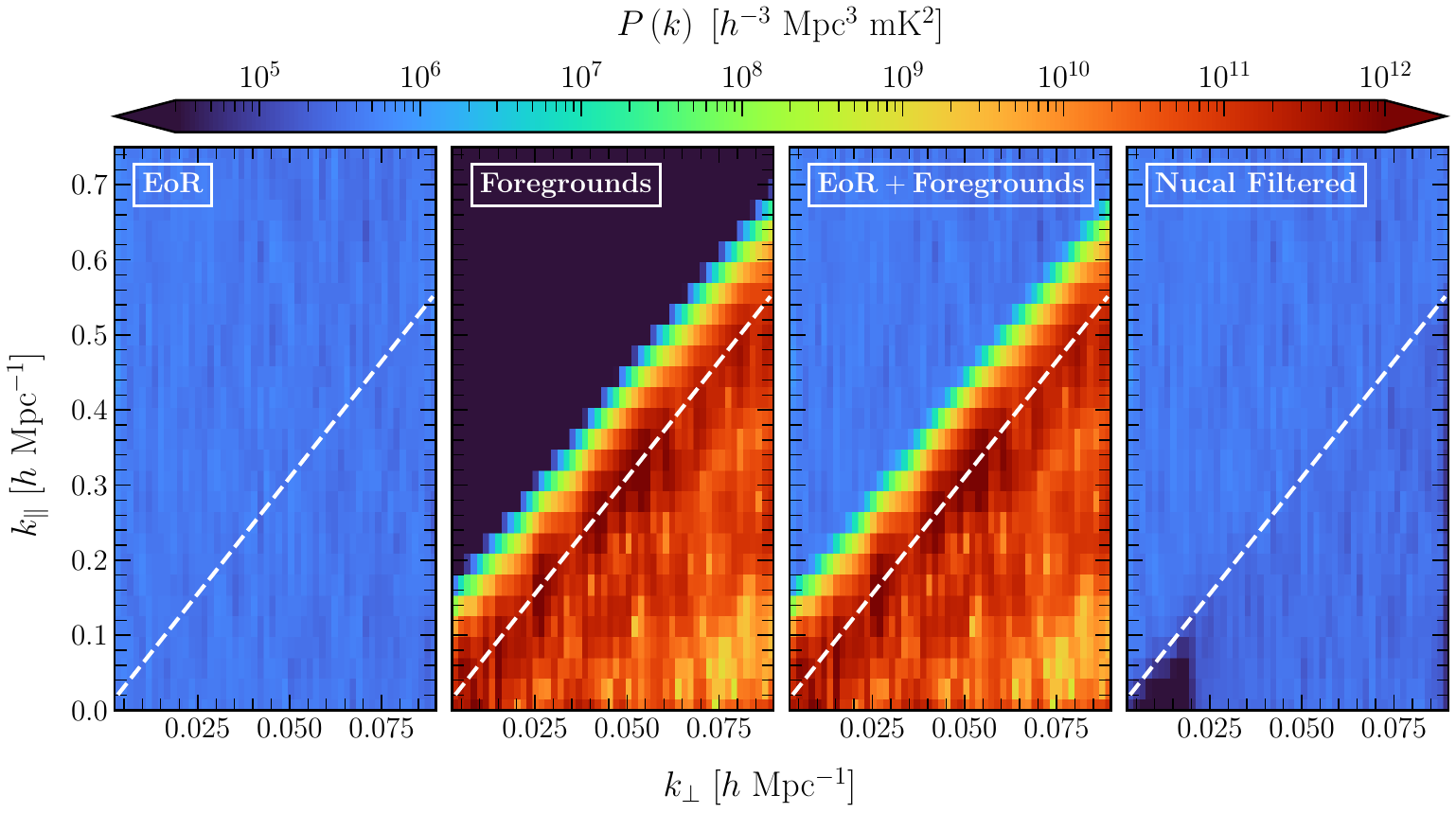}
\caption{Demonstration of the foreground filtering capabilities of spectral redundancy on a set of simulated visibilities from the most spectrally redundant group in the HERA array. Fitting a \nucal{} model suppresses foregrounds within the wedge, but is able to retain 21\,cm signal in simulated, noise-free simulations. By fitting versus $u$ and $\nu$, \nucal \ is able to remove foreground emission from a point-source model with varying spectral indices and a realistic level of beam chromaticity. In the first three subplots, we show the 2D power spectra of EoR, foregrounds, and the sum of both components. In the right-most panel, we plot the 2D power spectrum of the full simulated data (third panel) but with the \nucal{} model subtracted. By subtracting off our best fit model, we are able to substantially suppress foregrounds across a wide range of $k_{\perp}$ modes while retaining 21\,cm signal. Signal loss from overfitting appears in the last panel at low $k_{\parallel}$ and $k_{\perp}$ where spectral redundancy is minimal and 21\,cm modes are degenerate with \nucal{} foreground filters. We examine that loss more directly in \Cref{fig:signal_loss}.}
\label{fig:nucal_subtraction}
\end{figure*}

In \Cref{fig:nucal_subtraction}, we display the results of our foreground filtered visibilties as 2D power spectra for 21\,cm signal, foregrounds, foregrounds and 21\,cm and the filtered foregrounds and 21\,cm signal. We find that \nucal{} is able to substantially suppress foreground modes within the wedge, while revealing relatively clean EoR signal. We find substantial signal loss---i.e., incorporation of EoR fluctuations into the \nucal{} model, which is then subtracted off---at low $k_\parallel$ and $k_{\perp}$. This is clearest for short baselines, which have the least spectral redundancy (see \Cref{fig:tracks}). This lack of spectral redundancy leads to degeneracies between the foreground filters and 21\,cm signal that leads to an over-subtraction of cosmological signal within the wedge. Note that this result is for the single best-sampled spectrally-redundant group with 58 unique baselines, all oriented in the same direction (see \Cref{fig:antpos}); we should expect less effective subtraction for other groups in HERA. Many orientations within HERA have too few baselines to fit the smooth foreground degrees of freedom in \Cref{eq:NUCALMODEL}. Of the 61,075 total single-polarization baselines in HERA-350, which sample 6,610 unique baseline vectors, 26.8\% belong to baseline groups with 10 or more unique baselines---enough to be useful for \nucal{} (see \Cref{sec:nucal}). 13.5\% belong to highly-spectrally-redundant baseline groups with 25 or more unique baselines. 

\begin{figure}
\centering
\includegraphics[width=.99\columnwidth]{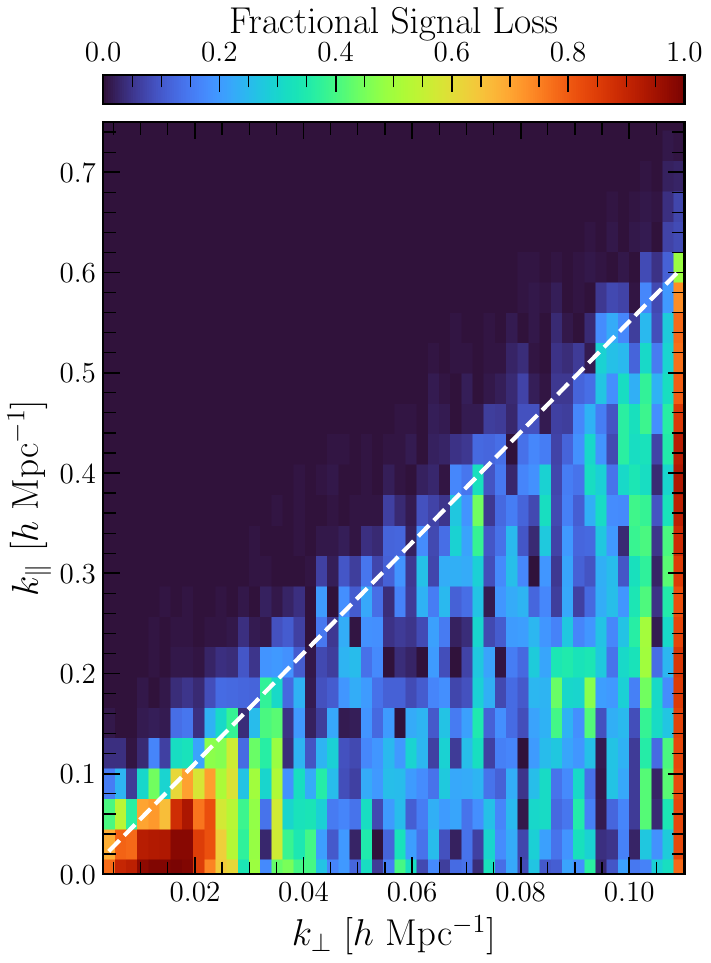}
\caption{Fractional signal loss computed following foreground filtering, shown in \Cref{fig:nucal_subtraction}. At low $k_{\perp}$ and low $k_{\parallel}$ baselines which contribute to these measurements of the power spectra have poor spectral-redundancy, leading to over-subtraction of the 21\,cm signal. At intermediate $k_{\perp}$ scales, more baselines overlap a given $uv$-mode, leading to better differentiation between foregrounds and 21\,cm signal and reduced ($\sim$$20\%$) signal loss within the foreground wedge. Signal loss is largely negligible in the EoR window, given that the DPSS modeling basis is restricted to producing wedge-like shapes in delay space. There is some signal beyond the horizon given by the dotted white line due to either the Blackman-Harris taper used or due to the small amount of chromaticity allowed by the DPSS basis functions used to model the frequency axis.}
\label{fig:signal_loss}
\end{figure}

In \Cref{fig:signal_loss}, we quantify the level of signal loss from overfitting during foreground subtraction by comparing the filtered power spectrum to the input EoR power spectrum. We find that some amount of signal is lost throughout the wedge and slightly beyond the horizon, due to the small amount of chromaticity given the DPSS basis modeling the frequency axis. Across the range of $k_{\perp}$ modes which are well-sampled, we find a signal loss value of $\sim$$20\%$. This is a non-negligible amount of signal, to be sure, but it is certainly preferable to losing 100\% of the EoR signal inside the wedge that one has to accept with a strategy of foreground avoidance. 

Such signal loss is not unexpected; the \nucal{} model fits any power at low $\eta$ regardless of its astrophysical origin. In a delay spectrum, that power gets spread across $\tau$ modes, which explains why we see signal loss throughout and even a bit beyond the wedge. However, the part of the power of this technique is the ability to use information from many other frequencies to better estimate (and thus remove) low $\eta$ structure, which is overwhelmingly foreground-dominated. While this technique suppresses the 21\,cm signal to a larger extent than other techniques \citep{2018MNRAS.478.3640M, 2022MNRAS.509.3693M}, the benefit of this approach is that it makes very few assumptions about the foregrounds and primary beam while still modeling them to good precision. 

Of course, to use this technique on real data, one would have to precisely and systematically model and account for the signal loss in observed spectrally redundant groups---a task beyond the scope of this work. Regardless, this level of foreground subtraction is a promising step towards mitigating the impact of foregrounds in certain array-types while retaining 21\,cm signal within the wedge. 

\section{Discussion} \label{sec:discussion}
In the previous sections, we explored the promise of utilizing spectral redundancy as both a method of performing accurate direction-independent gain calibration and foreground subtraction. Here, we discuss both the real-world limitations of and potential future extensions to the technique.

\subsection{Real-World Challenges of Leveraging Spectral Redundancy}  \label{subsec:realworld}

This work remains an incomplete study of what needs to be done to evaluate the ability of \nucal{} to calibrate real data and to remove foregrounds and recover 21\,cm signal within the wedge more generally. We have not fully explored important effects such as proper fixing of \nucal{}'s degenerate parameters, which requires reference to a sky-model, which are always incomplete. In the event that the sky-model significantly deviates from the observed visibility, improper degeneracy removal may impact the efficacy of foreground subtraction and complicate the recovery of cosmological modes within the wedge. Future work is needed to explore our ability to subtract foregrounds in the presence of sky-model incompleteness.

As with redundant calibration, spectrally redundant calibration requires that the frequency-dependent $uv$-plane be self-consistent between measurements made by different baselines. Several known effects can lead to deviations between baselines, which lead to non-redundancies in spectrally redundant calibration. These non-redundancies arise as a result of beam non-redundancies and antenna position errors, mutual coupling between elements in the array, and polarized emission.

The first is the most straightforward, as they are also non-redundancies associated with spatially redundant-baseline calibration: all antennas in the array must have the same primary beam response and be placed accurately. Spectral redundancy requires that beams are uniform across the array, otherwise baselines assumed to be spectrally redundant will measure different $uv$-planes, considering that the measured $uv$-plane is a convolution of the true $uv$-plane and the antenna beam. Attempting to jointly model baselines with \nucal{} that include antennas with non-redundant beams would likely introduce calibration errors into the antenna gains, complicating the recovery of 21\,cm signal in the EoR window. Deviations in both the primary beam size and beam pointing center have been shown to introduce spectral errors in redundant calibration \citep{2019MNRAS.487..537O}, and will likely be a point of concern when applying \nucal{} to data. \citet{2023ApJ...953..136K} show that beam errors can be mitigated in traditional redundant calibration by perform fringe rate filtering the data before calibrating, and could potentially help mitigate similar errors in spectrally redundant calibration. 

Similarly, antenna placement errors such as East-West/North-South positional errors or height offsets between elements may prove to introduce spurious structure in calibrated visibilities if one assumes that baselines sample the same modes in the $uv$-plane actually sample slightly different modes. \citep{2019MNRAS.487..537O} showed that gain errors introduce by antenna placement errors can be mitigated by down-weighting long baselines during calibration. However, the models produced by \textsc{nucal} are crucially informed by long baselines, and therefore down-weighting them may prove to be less effective for spectral redundancy than spatial redundancy. More work is needed to understand how to best handle antenna placement errors within the \textsc{nucal} framework.

In addition to beam non-redundancies and antenna placement errors, leakage of long baselines' visibilities into less chromatic short baselines due to mutual coupling between array elements is likely to pose a challenge to the convergence of \nucal{} in its current form. Given that these systematics extend significantly beyond the wedge, they are unable to be modeled with the set of DPSS basis functions used to model beam-sky product, and are unable to be removed via per-antenna gain calibration \citep{2019ApJ...884..105K, 2022MNRAS.514.1804J, 2023ApJ...945..124H, 2023MNRAS.522.1009C}. The inability to incorporate these effects into models of the foregrounds or gains will lead to chromatic gain errors if unaccounted for. However, one could imagine that \nucal{} could be modified to solve for first-order coupling coefficients, such as those presented in the framework developed in \cite{2022MNRAS.514.1804J}, in addition to the per-antenna gains, potentially reducing the impact of mutual coupling effects on the estimates of gain parameters.

The assumption that the foreground-beam product is spectrally compact may be violated by Faraday-rotated diffuse emission, particularly near the galactic center, where the effect is most prominent \citep{2009ApJ...702.1230T}. Spectral structure introduced by this polarized emission will cause variations at a fixed $u$ that we will be unable to model using our smooth set of basis functions, likely leading to chromatic errors in the gain estimates if that are not properly accounted for. One approach to calibrating visibilities which we expect to be highly polarized, is to require that the foregrounds in psuedo-Stokes I be smooth as a function of frequency rather than assuming that either of the linear polarizations are spectrally redundant. This formulation should help to reduce the impact of Faraday-rotated polarized emission on the estimates of the gains.

In addition to the non-idealities of real-world instruments and calibration, we did not provide a full study of the factors affecting signal loss when subtracting \nucal{} modeled foregrounds from simulated visibilities. Quantifying the precise level of signal loss is crucial when performing this kind of analysis, as the inclusion of wedge modes with substantial signal loss in an estimate of the power spectrum will artificially lower the amplitude of the estimated power spectrum \citep{2022ApJ...924...85A}. The exact amount of signal loss in a given wedge mode will depend on the precise set of basis vectors fit with \Cref{eq:NUCALMODEL}. It will also depend on the spectral smoothness of the instrument's primary beam response, the true spectral smoothness of the foregrounds, and the degree to which a given mode contributing to an estimate of power spectrum is sampled redundantly in frequency. These effects are fundamental to the array on which \nucal{} is being performed, and entirely independent of sky-model incompleteness and non-redundancy. The exploration of how variations in dish size, beam chromaticity, and antenna layout affect the level of signal loss within the wedge in 21\,cm measurements is also left for future work.

\subsection{Future Applications of \nucal}
While we have shown the benefits of spectral redundancy and spatio-spectral filtering for calibration and foregrounds subtraction, there are many more improvements of this technique and future applications which could make it more widely useful to the 21\,cm community.

Perhaps the most conceptually straightforward improvement to this technique is extending \nucal{} to solve for per-antenna gains. There are two potential ways in which this could be done. The first is by simply extending \Cref{eq:chi2_nucal} to incorporate per-antenna gains, instead of restricting the gain terms to the degenerate space of redundant-baseline calibration. While this approach is mathematically straightforward, extension to full per-antenna gains is extremely computationally challenging due to the massive increase in data volume and number of parameters required to solve for. Utilization of the \textsc{jax} environment to efficiently compute gradients and its ability to use GPUs to perform would likely help, but it is unclear if the approach would be computationally tractable for such as HERA without major improvements to the algorithm itself. Another complication of extending \nucal{} to per-antenna gains is the potential risk of introducing signal loss beyond the wedge. Because \nucal{} attempts to produce the smoothest possible calibrated visibilities, extending calibration to incorporate more degrees of freedom may lead to per-antenna gains overfitting 21\,cm signal, especially beyond the foreground wedge. More work is needed to understand how extending \nucal{} to incorporate more calibration degrees of freedom introduces the signal loss outside the wedge.

The second approach is slightly more limited, but more computationally efficient. Instead of solving for per-antennas gains and \nucal{} model parameters simultaneously, we could instead fit a \nucal{}-based model to data which have already been calibrated and then use that model as a sky-model to recompute per-antenna gains. Because the \nucal{} is highly constrained by the redundancy of an array and physically motivated to pick out foreground emission from sets of spectrally redundant baselines, it could be an excellent foreground model for sky-based calibration. One potential downside of this approach is that if the \nucal{} model is derived from visibilities with significant errors due to prior miscalibration and there is a lack of highly spectrally-redundant baselines within the array, this approach may be unlikely to fix those errors. However, both approaches are interesting avenues for further exploration and warrant future work.

As presented, \nucal{} is currently limited to modeling foregrounds in baselines which are part of groups that redundantly sample the same spatially Fourier modes at different frequencies. While some arrays do take advantage of redundant baseline measurements and therefore often redundantly sample the same $uv$-modes along radial spokes of the $uv$ plane, arrays which are not radially redundant will be unable to take advantage of the benefits of \nucal{} as currently formulated. However, given the flexibility of our modeling basis functions, \nucal's modeling \Cref{eq:NUCALMODEL} could in principle be modified such that the entire $uv$ plane be modeled jointly in a higher dimensional representation of these DPSS eigenfunctions. 

Doing so would significantly increase the number of free parameters incorporated into the foreground model, but will improve the generality of the approach to allow for the calibration of array types which do not explicitly sample the same $u$ modes. Additionally, this approach may improve the ability for non-radially redundant arrays to better model and subtract foregrounds by empirically estimating the cross-frequency covariance between baselines which samples nearby modes in the $uv$-plane without explicitly simulating foregrounds with a model of the sky and instrument primary beam. This approach to using modeling the entire frequency-dependent $uv$-plane may also offer a reduction in signal loss incurred by subtracting foregrounds with a \nucal{} model. Such an exploration and its implementation is outside the scope of this paper and left for future work.

Given the extreme dynamic range challenge 21\,cm cosmology presents, 21\,cm arrays must be co-designed to allow for an analysis which best allows for the separation of foregrounds from 21\,cm signal. Next-generation 21\,cm experiments will need to restore access to modes within the wedge in order to directly image cosmological hydrogen. Such three-dimensional images of neutral hydrogen support cross-correlation studies with other probes of large-scale structure, providing verification of a detection of the 21\,cm and unlocking stronger constraints on the physics of reionization \citep{2015ApJ...800..128B, 2017arXiv170909066K, 2023ApJ...944...59L}. Just as HERA's highly-regular array configuration is designed to prioritize spatial redundancy \citep{2016ApJ...826..181D, 2017PASP..129d5001D}, future array designs may be optimized for both spatial and spectral redundancy, allowing for calibration and foreground modeling with \nucal{} and potentially enabling for the recovery of cosmological modes within the foreground wedge that could be used for imaging. Future work is needed to investigate how new array configurations and element design can be leveraged to maximize the effectiveness of this technique.

\section{Summary}\label{sec:conclude}
In this paper, we introduced a novel technique for precisely calibrating interferometers for 21\,cm cosmology known as \emph{spectrally redundant calibration}, or \nucal{} for short. Inspired by traditional spatially redundant calibration, our approach solves for antenna gains by exploiting the correlations between baselines which redundantly sample the same angular Fourier modes in the $uv$-plane at different frequencies. Leveraging the fact that baselines which sample the same angular scales see highly correlated beam-sky products at different frequencies, we model this redundantly sampled $u-\nu$ plane as a linear combination of a highly-efficient set of smooth basis vectors known as discrete prolate spheroidal sequences (DPSS). 

We demonstrate in \Cref{sec:nucal} our ability to calibrate out arbitrary spectral structure in redundant calibration degeneracies in a set of simulated HERA visibilities and show that \nucal{} explicitly preserves the EoR window by preventing fine-scale calibration errors from coupling with foreground power. Because this approach relies on self-consistency between baselines at different frequencies, we do not need detailed models of the sky and beam to perform calibration while ensuring that our calibrated visibilities remain spectrally smooth.

While this technique does require reference to a sky model to solve for a set of additional parameters which are degenerate with our empirically-motivated foreground model, the number of degenerate parameters that are required to be solved for are substantially reduced compared to that of traditional redundant calibration and are limited to shapes residing within the foreground wedge. Assuming that the algorithm reaches convergence to the global minima, this significantly reduces the amount of spurious spectral structure that can be introduced via calibration error due to sky model incompleteness.

One of the most exciting aspects of this work is \nucal{}'s ability to precisely model and subtract foregrounds from baselines, which redundantly sample the same modes in the $uv$-plane. As shown in \Cref{sec:wedge_recovery}, combining spectral redundancy with spatio-spectral filtering is able to accurately model the beam-weighted sky even in case in which realistic chromatic beams and foregrounds are used. Assuming that the $uv$-sampling is dense enough to break the degeneracy between variation in the spatial and spectral axes, modeling foregrounds baselines in the same radial heading using the DPSS basis shows promise for recovering 21\,cm signal within the wedge. 

One must be cautious when applying this technique to real data, as overfitting can lead to cosmological signal loss, especially when a given $uv$-mode is not sufficiently redundantly sampled by the baselines in its unique orientation. As this effect is dependent on the assumed chromaticity of the sky and beam, the antenna layout of the array, and observing frequencies. Future work will attempt to quantify this signal loss for different array types. Despite those caveats and limitations, our technique is the first proposal to enable the estimation of 21\,cm signal within the wedge without detailed knowledge of foregrounds or the primary beam and is thus perhaps the first step toward recovering $\sim$80\% of the sensitivity that interferometers like HERA must sacrifice by avoiding the foreground wedge

\section*{Acknowledgements}
The authors gratefully acknowledge helpful discussions with Zachary Martinot and Mike Wilensky. This material is based upon work supported by the National Science Foundation under Grant \#1636646 and \#1836019 and institutional support from the HERA collaboration partners. This research is funded in part by the Gordon and Betty Moore Foundation. HERA is hosted by the South African Radio Astronomy Observatory, which is a facility of the National Research Foundation, an agency of the Department of Science and Innovation. J.S. Dillon gratefully acknowledges the support of the NSF AAPF award \#1701536.

\subsection*{Software}
This work was enabled by a number of software packages including \textsc{matplotlib}\footnote{\url{https://matplotlib.org/}} \citep{Hunter:2007}, \textsc{numpy}\footnote{\url{https://numpy.org/}} \citep{oliphant2006guide}, \textsc{scipy}\footnote{\url{https://www.scipy.org/}} \citep{2020SciPy-NMeth}, \textsc{jax}\footnote{\url{https://github.com/google/jax}} \citep{jax2018github} for data analysis and modeling, and \textsc{optax}\footnote{\url{https://github.com/google-deepmind/optax}} \citep{deepmind2020jax} for the gradient descent framework and library of optimizers.

\section*{Data Availability}
The data underlying this article (e.g. the visibility simulations) will be shared on reasonable request to the corresponding author.


\bibliography{references, software, non_ads_refs}
\bibliographystyle{aasjournal}

\appendix
\section{Nucal vs. Calamity}\label{appendix:calamity}
\nucal{} bears several similarities to the recently proposed \calamity \ technique \citep{2022ApJ...938..151E} due to its use of first-order optimization to determine gain solutions and the use of DPSS modeling. In this appendix, we highlight the specific similarities and differences between the two techniques by casting them both under the formalism adopted in the rest of the paper.  \calamity \ and \nucal \ differ primarily in two ways.

\subsection{Different foreground modeling functions}

Let the set of all frequencies sampled by every baseline be the same, which we denote with the vector ${\bf F}$ with length $N_F$. From \Cref{eq:NUCALMODEL}, we can describe any per-baseline modeling approach like the one described in \citet{2022ApJ...938..151E} with a set of $\phi_k$ basis functions where the vectors are partitioned into $b$ different sets where the $m^{th}$ set has length $n_m$ different functions, $\left\{ \varphi_{0,0}, ..., \varphi_{0, n_0}\right\}, \left\{\varphi_{1, 0}, ..., \varphi_{1, n_1} \right \}, \hdots, \left\{\varphi_{b, 0}, ..., \varphi_{b, n_b} \right \}$ and each $m^{\rm th}$ baseline is centered at the set of ${\bf u}$ values in the $N_F$ length set $\boldsymbol{\mathsf{U}}_m$. The first index of each function labels the baseline, and the second index labels the basis functions necessary to model that baseline.

\begin{equation}\label{eq:CALAMITY}
    \varphi_{m, n}({\bf u}, \nu_i) = \begin{cases}
    \phi_{m,n}(\nu) & {\bf u} \in \boldsymbol{\mathsf{U}}_m \text{ and } \nu \in {\bf F} \\ 
    0 & \text{otherwise}
    \end{cases}
\end{equation}

In the per-baseline modeling case, we work with a set of basis functions that only having support over a single baseline. In practice, we fit discrete vectors to our data the basis vectors in the per-baseline modeling case are non-zero over a single baseline. For the DPSS modeling used in \calamity, we choose $\phi_{m,n}$ equal to 

\begin{equation}
    \phi_{m,n} = u_n(\nu, \mathcal{W}_m)
\end{equation}

where $u_n$ is the $n^{\rm th}$ Slepian sequence \citep{1978ATTTJ..57.1371S} with normalized bandwidth of $\mathcal{W}_m = 2 B \tau_m$ where $\tau_m$ is the delay width that is chosen to describe the foregrounds modulated by the primary beam.

\subsection{Different gain modeling assumptions}
In \nucal, we assume that the redundant degrees of freedom in the gains have already been removed through some redundant calibration strategy. All that is left to solve for in the gains is a common per-frequency absolute gain factor $A(\nu)$ and the tip-tilt term $\Phi(\nu)$. In \calamity, we retain and fit per-frequency complex gain for every antenna in the array. This limits the utility of \nucal{} as specifically laid out in this paper to redundant arrays with identical beams while \calamity{} can, in principle, be applied to arbitrary array layouts with non-redundant beams. Reducing the gain degrees of freedom may reduce the amount of signal loss suffered by our measurement for a fixed array layout, though the larger number of independent $uv$ modes sampled by non-redundant arrays could counteract this. We leave a detailed exploration of the tradeoffs between gain modeling assumptions and array layout to future work.

In summary, \nucal{} reduces to \calamity{} in the limit that the antenna gains are described by a tip-tilt parameter and an amplitude degeneracy, the true visibility modeling functions have support only over individual baselines, and are described by discrete Slepian sequences instead of sampled prolate spheroidal wave functions.
\end{document}